\documentclass[prb,twocolumn,amsmath,amssymb,showpacs]{revtex4}
\usepackage{graphicx}

\begin{document}

%
%

\title{Inhomogeneous Charge Textures Stabilized by Electron-Phonon 
Interactions in the t-J Model}

\author{Jos\'e Riera$^{1}$ and Adriana Moreo$^{2}$}

\affiliation{$^1$ Instituto de F\'isica Rosario, Consejo Nacional de 
\\Investigaciones Cient\'ificas y T\'ecnicas, y Departmento de F\'isica,
\\Universidad Nacional de Rosario, Avenida Pellegrini 250, 2000-Rosario, 
Argentina}

\affiliation{$^2$Department of Physics and Astronomy,University of Tennessee,
Knoxville, TN 37966-1200 and 
\\Condensed Matter Sciences Division, 
Oak Ridge National Laboratory,Oak Ridge, 
TN 37831-6032 }

\date{\today}

\begin{abstract}

We study the effect of diagonal and off-diagonal electron-phonon coupling in 
the ground state properties of the t-J model. Adiabatic and quantum phonons 
are considered using Lanczos techniques. Charge tiles and 
stripe phases with mobile holes (localized holes) are observed at 
intermediate (large) 
values of the diagonal 
electron-phonon coupling. The stripes are stabilized by half-breathing modes, 
while the tiles arise due to the development of extended breathing modes.
Off-diagonal terms destabilize the charge inhomogeneous structures with 
mobile holes by renormalizing the diagonal coupling but do not produce new 
phases. 
Buckling modes are also studied and they seem to induce a gradual 
phase separation 
between hole rich and hole poor regions. 
The pairing correlations are strongly suppressed when the holes are localized.
However, in charge inhomogeneous states with mobile holes 
no dramatic changes, compared with the uniform state, are observed in the 
pairing correlations indicating that D-wave 
pairing and moderate electron-phonon interactions can coexist.

\end{abstract}


\maketitle

\section{Introduction}

Electron-phonon interactions (EPI) are at the heart of the pairing 
mechanism in the BCS theory \cite{BCS} that successfully describes traditional 
superconductors. It is believed, however, that EPI cannot produce the high 
critical temperatures observed in cuprate superconductors and, for this reason,
most of the research in the field has focused on the interaction between 
charge and magnetic degrees of freedom as a source of the pairing mechanism 
in the cuprates.\cite{magnetic,review} 

Despite almost 20 years of intensive research, the 
pairing mechanism in the cuprates is still unknown. The experimental 
evidence indicating that phonons and lattice degrees of freedom play an active 
role in these materials keeps mounting.\cite{Bianconi,Egami,Lanzara,Tranquada}
In addition, inhomogeneous structures described as stripes, patches, tiles, 
etc., have been detected in the underdoped regime of several 
cuprates.\cite{Tranquada,Davis,Yazdani}  The emerging phase complexity is 
reminiscent of the experimental data for manganites where competing 
electronic, magnetic and lattice degrees of freedom are responsible for the 
rich phase structure.\cite{manrev}
It is also likely that the strong electronic correlations present in these 
materials, in particular the proximity to a Mott insulating phase in the 
cuprates, could radically change the effects of EPI with respect to those
in conventional metals and superconductors.

Most theoretical efforts that tried to incorporate EPI in models for the 
cuprates were performed in the 90's and they focused on whether the 
interactions could produce D-wave pairing 
(instead of S-wave)\cite{nazarenko,anet},
and on the tendency of phonons towards the stabilization of charge density 
wave (CDW) states that could compete with superconductivity (SC). In order to 
observe CDW states most of these calculations were performed at quarter 
filling.\cite{jose,doug} Currently, our aim is to understand whether
EPI stabilize or destabilize charge
stripes, and what kind of inhomogeneous textures, if any, develop. Instead of 
focusing on the phonons as an alternative source of the pairing mechanism 
we want to 
explore whether the interplay of magnetic and phonon interactions with
the electrons may lead to an enhancement of the pairing already
observed in purely electronic models of high T$_c$ superconductors.\cite{bili}

In the early studies only diagonal EPI were 
considered and the effects on hopping terms and Heisenberg coupling due to the 
lattice distortions (off-diagonal terms) were disregarded. Most of the 
numerical work was performed with on-site Holstein phonons, and when more 
extended modes were considered the focus was on breathing modes\cite{doug} 
that would tend to stabilize CDW states and in buckling modes that were known 
to produce the tilting of the octahedra in YBCO.\cite{nazarenko} At present, 
the experimental evidence indicates that the electron-phonon coupling to the 
breathing mode is strongly anisotropic and, as a result, 
half-breathing modes seem to have the 
strongest EPI in the high Tc cuprates.\cite{Egami,Tranquada} In addition, some 
authors believe that off-diagonal couplings could be stronger than the 
diagonal ones.\cite{ishihara}  

Most of the recent studies of EPI in models for the cuprates have been 
performed using mean-field, slave-boson, or LDA 
approximations.\cite{Olle,Dev,bishop,ishihara} Here, we propose to revisit 
the t-J model and study the effects of EPI with unbiased numerical techniques.
In order to find the relevant phononic modes we will start by studying 
adiabatic phonons, i.e. $\omega=0$. Results at finite frequency will be 
presented for the most relevant modes only.

The paper is organized as follows: Section II describes the study of 
adiabatic phonons. Diagonal coupling to breathing modes is discussed in 
subsection A while subsection B contains results for diagonal coupling to 
buckling modes. The effect of off-diagonal couplings in both cases is 
discussed in subsection C. Quantum phonons for the physically
important case of half-breathing and buckling modes are discussed in 
Section III. Conclusions and final remarks are the subject of Section IV.

\section{Adiabatic Phonons}

\subsection{Breathing modes}

The Hamiltonian for the t-J model with adiabatic phonons is given by
$$
{\rm H=-\sum_{\langle{\bf ij}\rangle\sigma}t_{\bf ij}
({\tilde c}^{\dagger}_{{\bf i}\sigma}
{\tilde c}_{{\bf j}\sigma}+h.c.)} 
+\sum_{\langle {\bf ij} \rangle}
J_{\bf ij}{\bf{S}}_{\bf i} \cdot{\bf{S}_{\bf j}}+H_{e-ph}+H_{ph},
\eqno(1a)
$$
\noindent with
$$
H_{e-ph}=\lambda_0\sum_{{\bf i}}u({\bf i}) n_{{\bf i}},
\eqno(1b)
$$
\noindent and
$$
H_{ph}={\kappa\over{2}}\sum_{{\bf i},\mu}(u_{{\bf i},\mu})^2,
\eqno(1c)
$$
where
$$
u({\bf i})=u_{{\bf i}, x}-u_{{\bf i-\hat x}, x}+
u_{{\bf i}, y}-u_{{\bf i-\hat y}, y},
\eqno(2)
$$
$$
t_{{\bf i,j}}=t\{1+\lambda_t[u({\bf i})+u({\bf j})]\},
\eqno(3)
$$
\noindent and
 $$
J_{{\bf i,j}}=J\{1+\lambda_J[u({\bf i})+u({\bf j})]\}.
\eqno(4)
$$

${\rm {\tilde c}^{\dagger}_{{\bf i}\sigma} }$ creates an electron
at site ${\bf i}=({\rm i_x,i_y})$ of a two dimensional square lattice 
with spin projection $\sigma$; 
${\bf{S}_i}$ is the spin-${1\over{2}}$ operator at site ${\bf i}$,
${ \langle {\bf ij} \rangle }$ denotes nearest-neighbor
lattice sites,
${\rm t}$ is the hopping amplitude, and ${\rm J>0}$ is the
antiferromagnetic exchange coupling.
In the following, $t=1$ and $J=0.4$ will be adopted unless otherwise
stated.
Doubly occupancy is not allowed in this model.
$n_{{\bf i}}=
\sum_{\sigma}c^{\dagger}_{{\bf i}\sigma}
c_{{\bf i}\sigma}$ is the electronic density on site ${\bf i}$. $\lambda_0$ is 
the diagonal electron-phonon coupling and $\kappa$ is the stiffness constant. 
$\lambda_t$ and $\lambda_J$ are off-diagonal coupling constants for the 
hopping and Heisenberg interactions respectively.
The lattice
distortions $u_{{\bf i},\mu}$  measure the displacement along the
directions $\mu=\hat x$ or $\hat y$ of oxygen ions located at the center
of the lattice's links in the equilibrium position, i.e.,
$u_{{\bf i},\mu}=0$.
Ground state properties of this model will be studied
using the Lanczos method\cite{review} on  square clusters with periodic 
boundary 
conditions (PBC). The results will be presented on 
$4\times 4$ clusters. Despite the small size we want to remark that all 
previous exact numerical 
studies were carried out on smaller clusters, $\sqrt{8}\times\sqrt{8}$
in Ref.\onlinecite{doug} and  $\sqrt{10}\times\sqrt{10}$ in 
Ref.\onlinecite{jose}. 
In addition, larger tilted clusters do not possess the symmetry to fit 
horizontal or vertical stripes. Notice that the ionic displacements 
$u_{{\bf i},\mu}$ will be obtained by finding the configuration that 
minimizes the total energy\cite{jose1} and without imposing {\it a priori} a
given pattern for them. Details of the procedure were 
presented in Ref.\onlinecite{jose1} and will not be repeated here.

\vskip2cm
\begin{figure}[thbp]
\begin{center}
\includegraphics[width=6cm,angle=0]{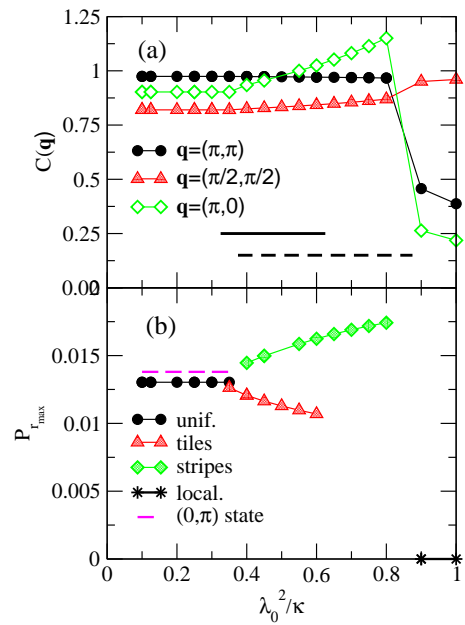}
\vskip 0.3cm
\caption{Results for the diagonal breathing mode for four holes in a 
$4\times 4$ lattice and $J=0.4$. a) Charge structure factor 
for different values of the momentum {\bf q}
as a function of the diagonal electron-phonon coupling. 
The full (dashed) line indicates 
the region where the tile (stripe) ground state is stable; b) Pairing 
correlations. The different symbols represent
different ground states described in the text.}
\label{breadiagh4}
\end{center}
\end{figure}
 
\begin{figure}[thbp]
\begin{center}
\includegraphics[width=4cm,angle=0]{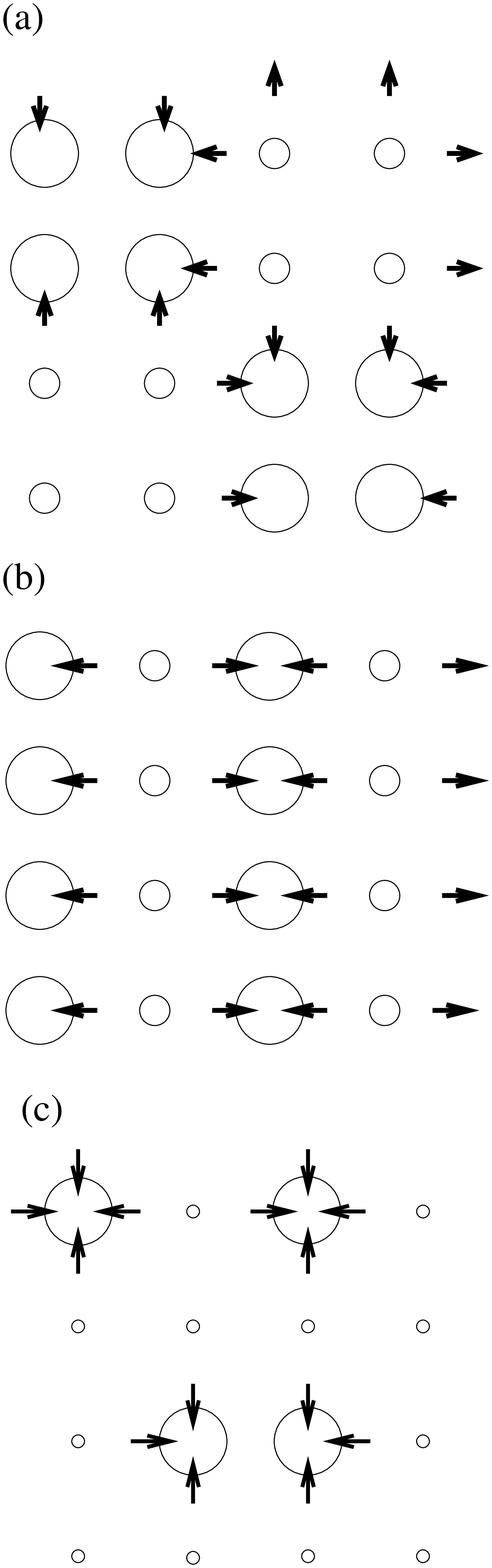}
\hspace{0.25cm}
\caption{(a)  Schematic representation of the tile structure stabilized by 
extended breathing phononic modes; the radius of 
the circles is proportional to the hole density $x_i=1-n_i$ and the arrows 
represent the ionic displacements; (b) Schematic representation of the charge 
stripes stabilized by horizontal half-breathing phononic modes;  
(c) Schematic representation of the 
localized holes that result from large values of the diagonal electron-phonon 
coupling.}
\label{schembreah4}
\end{center}
\end{figure}

We will start by discussing the effects of the diagonal coupling, i.e., 
$\lambda_t=\lambda_J=0$ in Eqs. (3) and (4).
For the case of four holes, i.e., $\langle n\rangle=0.75$ 
in a $4 \times 4$ lattice, we have observed that the 
ground state presents a uniform charge distribution up to 
$\lambda_0^2/\kappa=0.3$. The probability of hole occupation 
$\langle n_h({\bf i})\rangle$ is the same for every site ${\bf i}$ and
the charge structure factor $C({\bf q})$, 
shown in Fig.~\ref{breadiagh4}(a), does not have a sharp peak for any value 
of the momentum ${\bf q}$. 
Notice that the uniform ground state in the $4\times 4$ lattice 
is three-fold degenerate, i.e. 
states with momentum
${\bf q}=(0,0)$, $(\pi,0)$ and $(0,\pi)$ have the same energy. The 
pairing\cite{foot1} 
correlations are not the same in these three degenerate uniform states.
In Fig.~\ref{breadiagh4}(b)
the filled black circles indicate the averaged pairing 
correlations at the 
maximum distance along the diagonal.
The pairing along the maximum diagonal distance for the state 
with momentum $(0,\pi)$ is indicated by the dashed line.
For $\lambda_0^2/\kappa\ge 0.3$ the uniform ground state is replaced by a 
charge inhomogeneous state with a tiled structure schematically 
displayed in Fig.~\ref{schembreah4}(a). The size of the circles is
proportional to the probability of 
finding a hole in the corresponding site. The large circles indicate a 
probability of around 0.3 while the smaller ones correspond to about 0.2, 
which indicates that it is more likely to find the four holes in the tiles
defined by the big circles than outside them. The state is stabilized by 
``extended breathing'' modes as denoted by the arrows displayed in 
Fig.~\ref{schembreah4}(a), which are proportional to the 
ionic displacements $u_{{\bf i},\mu}$. 
 Although we are not aware of breathing or extended breathing modes 
being mentioned as relevant in the cuprates it is worth mentioning that so 
called ``tiled'' structures have been observed in STM experiments on 
$\rm Na_xCa_{2-x}CuO_2Cl_2$.\cite{seamus} 
In addition, all efforts to stabilize the 
observed tiled state in the t-J model without phonons have been 
unfruitful.\cite{js} The pairing correlations  in this state, denoted by the 
triangles in Fig.~\ref{breadiagh4}(b), are slightly suppressed but not 
destroyed. 
This state remains stable up to $\lambda_0^2/\kappa=0.62$ but from 
 $\lambda_0^2/\kappa\ge 0.38$ it coexists with a stripe state schematically 
shown in Fig.~\ref{schembreah4}(b). As in the previous case, the
large (small) circles indicate a site hole density of 0.3 (0.2) in the 
figure. 

In agreement with neutron 
scattering results for the cuprates, four holes in the $4\times 4$ 
lattice produce
two stripes as opposed to only one.\cite{Tranquada} Also, in agreement with 
experimental data \cite{Egami}, we observed that vertical (horizontal) 
stripes are stabilized by half-breathing modes along $x$ ($y$). The ionic 
displacements for vertical stripes are presented in Fig.~\ref{schembreah4}(b). 
The stripe state has a maximum in the charge structure factor at momentum 
${\bf q}=(\pi,0)$ (vertical stripes) or $(0,\pi)$ (horizontal stripes) which 
can be observed in Fig.~\ref{breadiagh4}(a). An 
interesting characteristic of the stripe state is that the pairing 
correlations at the maximum diagonal distance (filled diamonds in
Fig.~\ref{breadiagh4}(b)) slightly increase when compared with the 
result obtained from the average of the three uniform states 
(filled circles). This result is important because it indicates that 
D-wave pairing correlations can survive in stripe-like states as long as the 
holes are mobile. 

When the stripe-state is stabilized 
the triple degeneracy in momentum of the ground state is broken. 
The ground state is now a singlet with momentum $(0,\pi)$ ($(\pi,0)$) for 
vertical (horizontal) stripes. The observed increase in the pairing 
correlations does not seem to arise due to the break down of the momentum
degeneracy since the pairing correlations in the uniform state with momentum
$(0,\pi)$ (dashed line in Fig.~1(b)) are weaker than in the stripe state. 
We have
also observed that in the stripe state the pairing correlations are stronger 
in the direction perpendicular to the stripes than in the direction parallel
to them.\cite{yucel}

The stripe state becomes the only ground state for 
$0.6\le\lambda_0^2/\kappa\le0.8$. The magnetic structure factor $S(q)$ in the 
stripe state has a peak at $(\pi/2,\pi)$ which indicates a $\pi$-shift. 

It is important to remark that a striped ground state has been very difficult 
to stabilize in the t-J model without using external fields or special 
boundary conditions.\cite{SW} However, our results show that 
a moderate diagonal electron-phonon coupling to breathing phononic 
modes stabilizes them in a wide region of parameter space. 
Local lattice distortions in the direction 
perpendicular to the stripe, i.e., half-breathing modes,  
are responsible for this. Coincidentally, these are the modes that most 
strongly couple to electrons in the 
cuprates.\cite{Egami} 

For $\lambda_0^2/\kappa> 0.8$ a state with localized 
holes, with one example schematically shown in Fig.~2(c), becomes the 
ground state. The holes are 
localized by displacements of the neighboring ions as seen in the figure. The
probability of finding a hole in the sites with the large (small) circles 
is almost 1 (0), which indicates that the holes are localized. The pairing 
correlations at long distance vanish in this state and they are indicated by 
the star symbols in Fig.~1(b).

\begin{figure}[thbp]
\begin{center}
\includegraphics[width=6cm,angle=0]{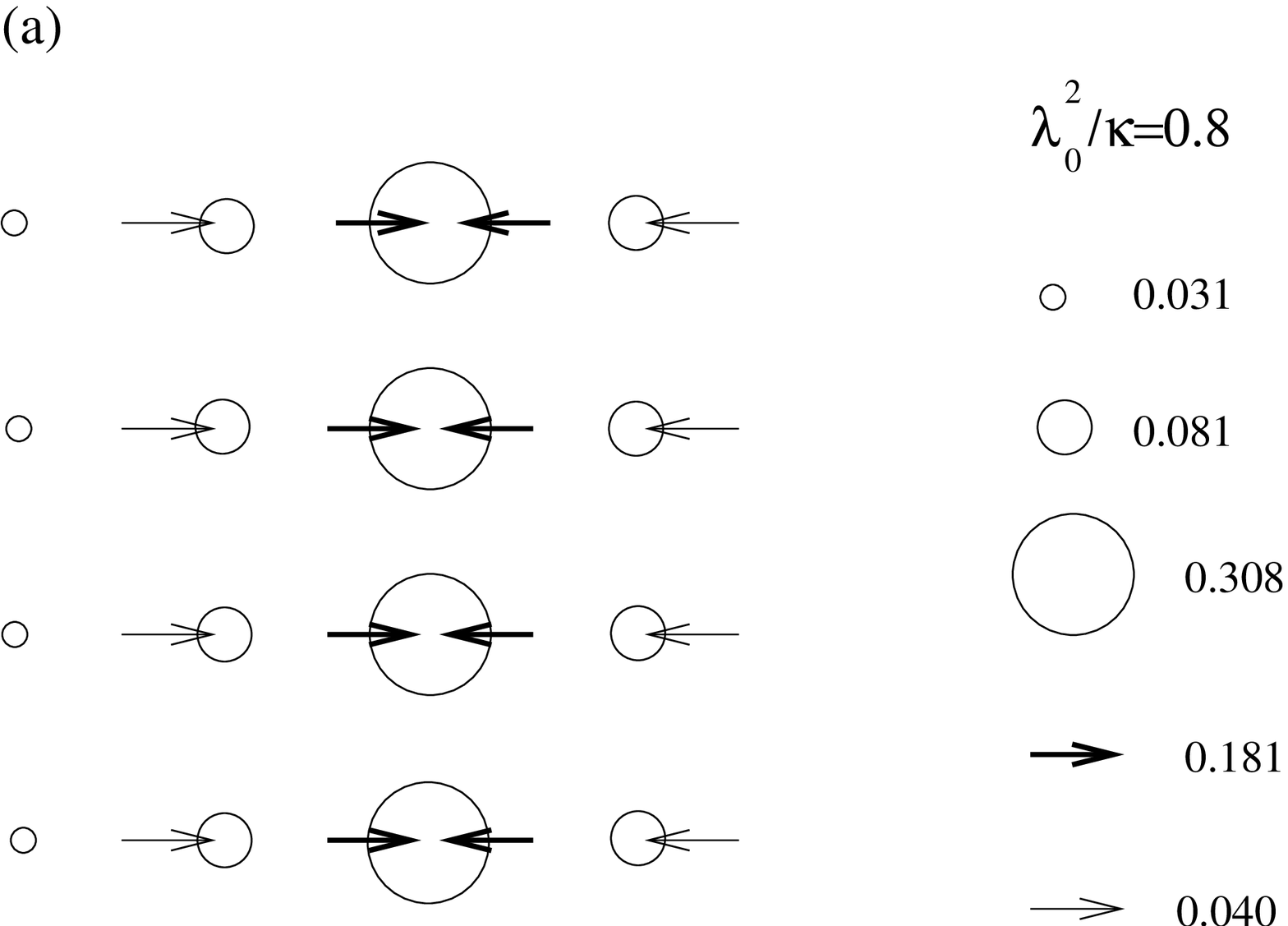}
\vskip 1cm
\includegraphics[width=6cm,angle=0]{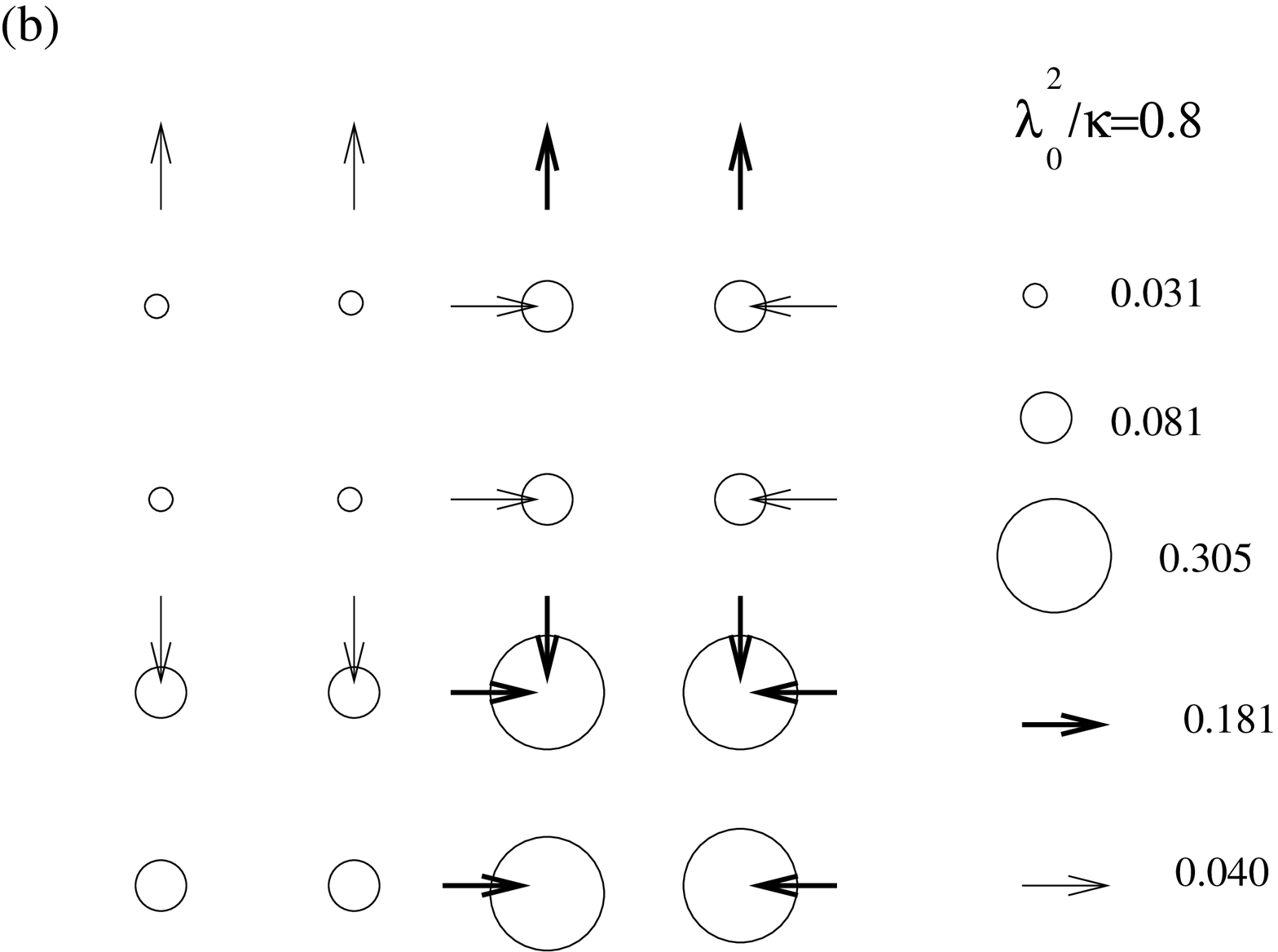}
\vskip 0.3cm
\caption{(a) Schematic representation of the charge 
stripes stabilized by horizontal half-breathing phononic modes for 2 holes in 
a $4\times 4$ lattice; the radius of the circles is proportional to the 
density of holes and the thickness of the arrows is proportional to the 
magnitude of the ionic displacements. 
(b) Same as (a) but for the tile structure stabilized by 
extended breathing phononic modes.}
\label{schembreah2}
\end{center}
\end{figure}

For two holes ($\langle n\rangle=0.875$) similar inhomogeneous charge 
structures were observed.
As for the case of four holes, stripe states, shown  in 
Fig.~\ref{schembreah2}(a), 
are stabilized by half-breathing phononic modes. As before, the size 
of the circles is proportional to the probability of finding a hole in the 
corresponding site. Two holes induce one stripe in the $4\times 4$ lattice
in agreement with neutron scattering results.\cite{Tranquada} 
However, notice the different 
thickness of the arrows in Fig.~\ref{schembreah2}(a) showing that the ionic
displacements are 
stronger next to the holes. This state has a complicated structure in momentum 
space and the charge structure factors present peaks at $(\pi/2,0)$, 
$(\pi,0)$ and $(3\pi/2,0)$.    
A tile structure with competing energy is also present in this case. The 
corresponding structure is shown in Fig.~\ref{schembreah2}(b). 
Notice that although 
it is more likely to find the holes in the tile, the holes are not localized 
since there is a finite probability of finding them outside. 
The tendency towards pairing is clear because it is more likely that the two 
holes will be found along the stripe or inside the plaquette.
Fully localized holes are 
observed for larger values of the diagonal electron-phonon coupling.

Qualitative changes are observed when 8 holes are introduced, i.e. at quarter 
filling. In agreement with earlier studies\cite{doug,jose} we found that
the phonon mode that 
leads to the ground state in this case is the full-breathing mode that 
stabilizes a checkerboard
charge density wave  state, as shown in Fig.~4(a) 
for $J=0.4$. 
The probability of finding holes associated to the size of the circles 
indicates that the holes are localized. The transition from the uniform state 
to the CDW state as a function of the diagonal coupling is characterized by a 
rapid increase of a peak at momentum $(\pi,\pi)$ in the charge structure 
factor, shown in Fig.~5(a), and for the development of incommensurability, 
along the diagonal, in the magnetic structure factor (see Fig.~5(b)).

We have observed that for larger values of $J$,  
a dimer state, shown in Fig.~4(b) for $J=1$, develops between the uniform 
and the CDW states. The new phase is clearly observed by monitoring the behavior 
of the charge structure factor, shown in Fig.~5(c) for $J=1$. A peak at momentum 
$(\pi/2,\pi)$ develops because, as it can be seen in Fig.~4(b), the dimer state 
breaks rotational invariance. At the same time a peak at momentum $(\pi,0)$ 
develops in the magnetic structure factor (see Fig.~5(d)).
The dimer state is 
stabilized by extended breathing phonon modes for intermediate values of 
the EPI, i.e.,
$0.5\le\lambda_0^2/\kappa\le 0.9$, for $J=1$. The dimers are localized pairs. 
This state 
was not observed in previous studies which considered only local breathing 
modes.\cite{doug,jose} This result is not surprising since it is well known 
that the $t-J$ model presents phase separation at $J/t\ge 3$ in the absence
of EPI. Thus, the proximity to phase separation causes the formation of 
dimers at moderated values of $\lambda_0^2/\kappa$ as $J$ increases. For a 
fixed $J$ the checkerboard CDW state is stabilized at sufficiently large 
values of the EPI.  

The pairing correlations
in the CDW and dimer states vanish at large distances which is not unexpected 
since in both cases the holes are localized. 
 

\begin{figure}[thbp]
\begin{center}
\includegraphics[width=6cm,angle=0]{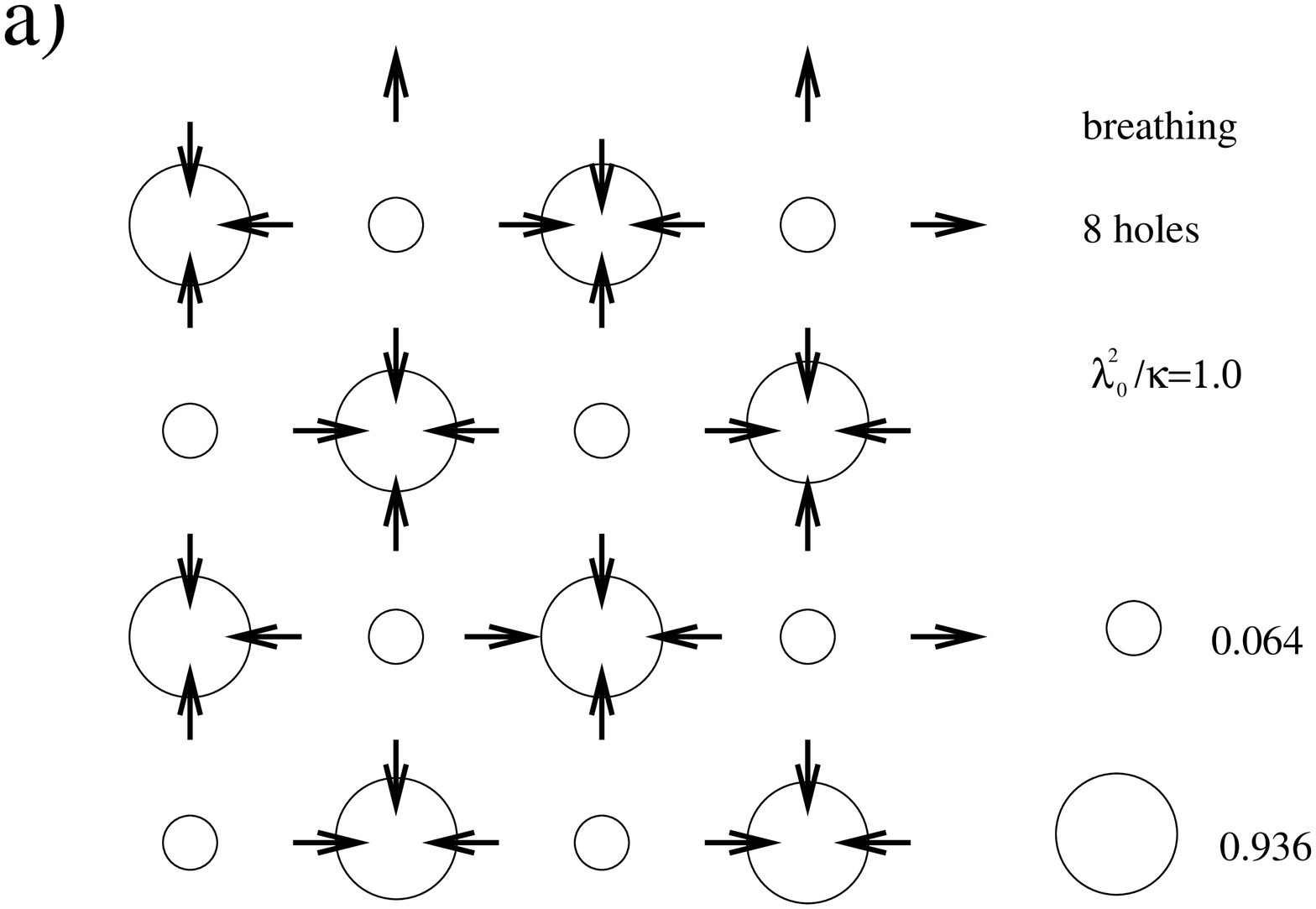}
\vskip 1cm
\includegraphics[width=6cm,angle=0]{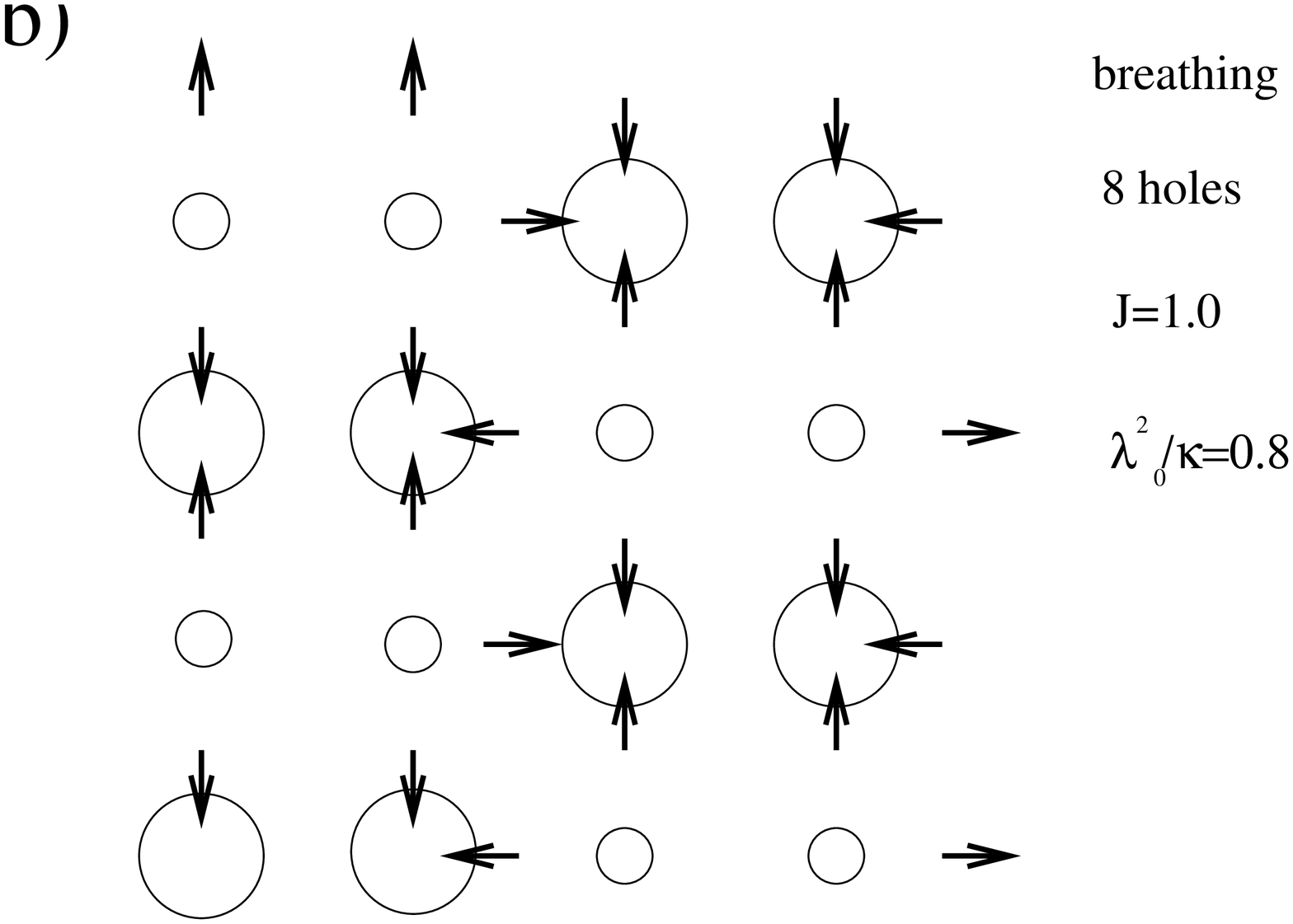}
\vskip 0.3cm
\caption{(a) Schematic representation of the CDW state 
stabilized by breathing phononic modes using 8 holes in 
a $4\times 4$ lattice for $J=0.4$; 
(b) Schematic representation of the dimer state stabilized by extended 
breathing
phononic modes at quarter filling (8 holes) for $J=1$.}
\label{fig4}
\end{center}
\end{figure}

\begin{figure}[thbp]
\begin{center}
\includegraphics[width=6cm,angle=0]{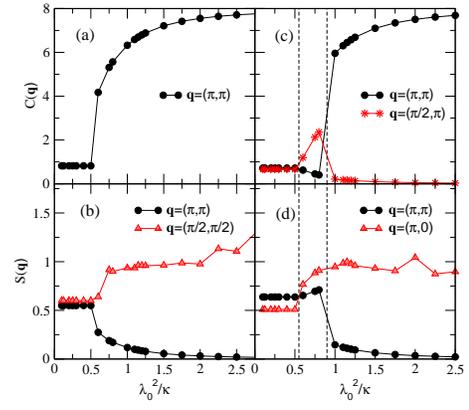}
\vskip 0.3cm
\caption{(a) Charge structure factor for 8 holes in 
a $4\times 4$ lattice for $J=0.4$; 
(b) Magnetic structure factor for the same parameters as (a); 
(c) Same as (a) for$J=1$; (d) same as (b) for $J=1$.}
\label{fig5}
\end{center}
\end{figure}

In summary, the previous results show that the diagonal electron-phonon 
coupling to breathing modes stabilizes charge inhomogeneous structures, 
some of which, like stripes and tiles, have been observed in the cuprates.
The pairing correlations are depressed when the holes 
are localized. However, in charge inhomogeneous states with {\it mobile} holes
the D-wave pairing correlations survive. Although it is not possible to
infer that long range correlations will develop in larger lattices, these 
results indicate that the interaction between magnetic and phononic degrees of
freedom does not necessarily destroy pairing.

\subsection{Buckling modes}

For completeness, the effect of buckling modes will also be considered.
Buckling modes are believed to be important in YBCO \cite{anet,normand}
and it was suggested that in combination with the magnetic interactions they 
could produce D-wave electronic pairing.\cite{nazarenko} The buckling mode is 
incorporated by replacing $u({\bf i})$ in Eq.(1) by
$$
u({\bf i})=u_{{\bf i}, x}+u_{{\bf i-\hat x}, x}+
u_{{\bf i}, y}+u_{{\bf i-\hat y}, y},
\eqno(5)
$$
\noindent while $t_{{\bf i,j}}$ and $J_{{\bf i,j}}$ have to be replaced by 
$$
t_{{\bf i,i+\mu}}=t[1+\lambda_t u_{{\bf i},\mu}],
\eqno(6)
$$
\noindent and
 $$
J_{{\bf i,i+\mu}}=J[1+\lambda_J u_{{\bf i},\mu}].
\eqno(7)
$$
\begin{figure}[thbp]
\begin{center}
\includegraphics[width=6cm,angle=0]{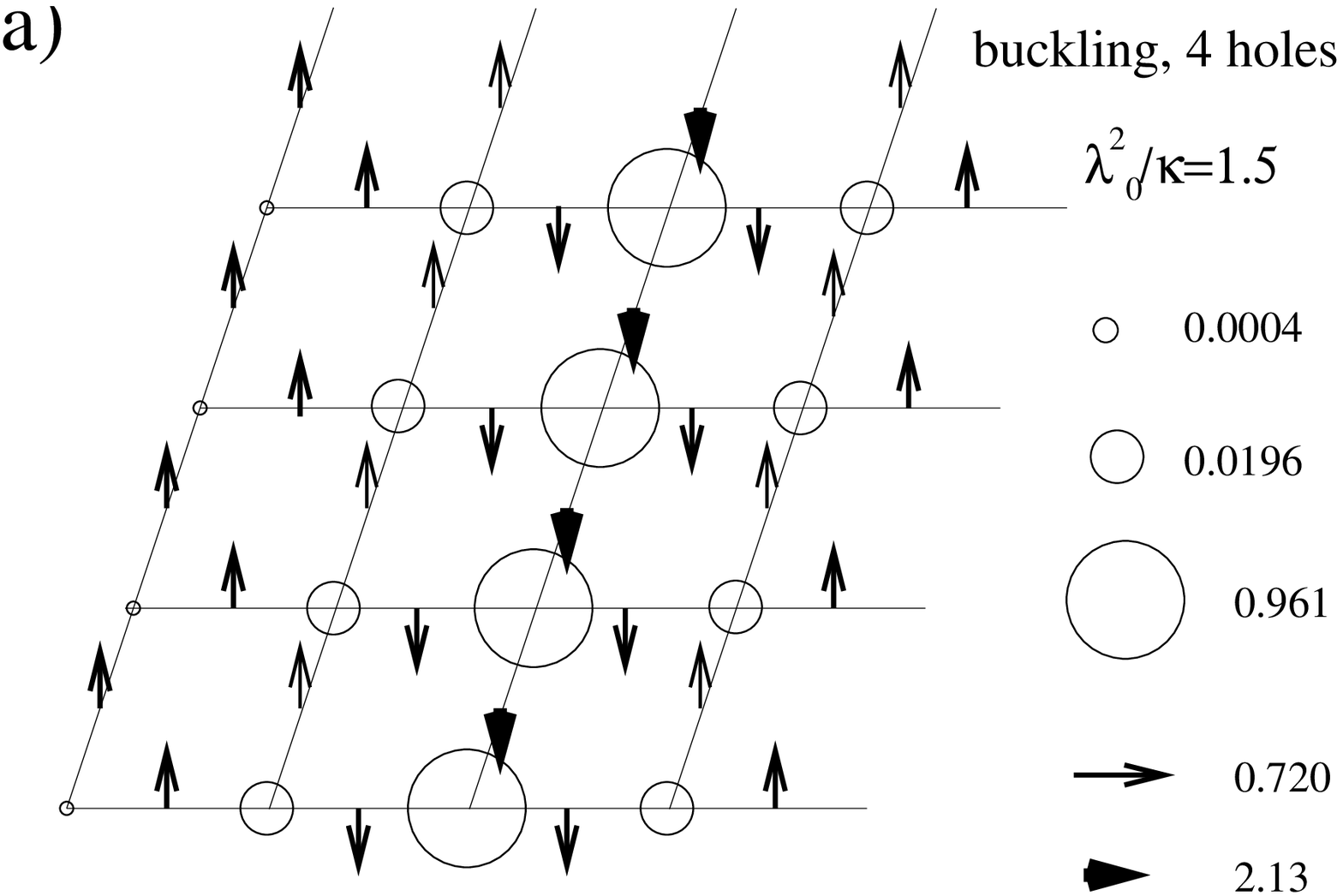}
\vskip 1cm
\includegraphics[width=6cm,angle=0]{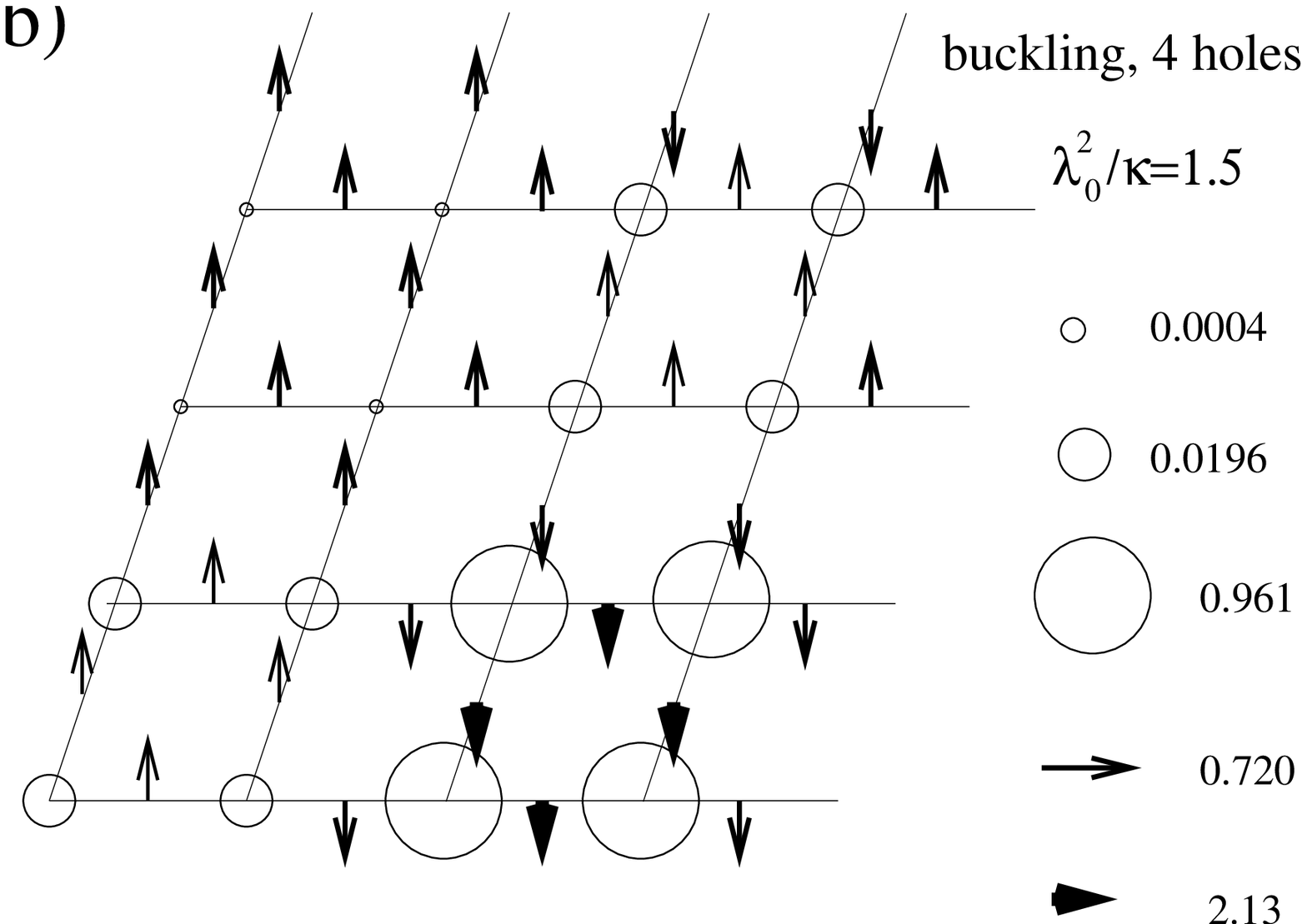}
\vskip 0.3cm
\caption{Schematic representation of the phase separated states 
induced by buckling phononic modes for 4 holes in 
a $4\times 4$ lattice: a) charge 
stripes; (b) the tile structure.}
\label{fig6}
\end{center}
\end{figure}
 
The most remarkable effects of adiabatic buckling modes are the following: 
1) stripe and tile states qualitatively similar to the ones observed with 
half-breathing modes are stabilized in a very narrow region 
at intermediate values of the diagonal coupling, and 2) the ground state 
phase separates for large values of $\lambda_0^2/\kappa$.
 
We have observed the phase separated state by doping with
 2, 4 and 8 holes. In the case of 2 holes pair localization 
appears. However, for 4 holes it can be seen that the holes localize in 
clusters 
instead of remaining  isolated as it was the case with breathing modes 
(see Fig.~2(c)). 
Examples of the phase separated clusters for 4 holes are shown in 
Fig.~6. The stripe (tile) structure in Fig.~6(a) (Fig.~6(b)) 
resembles the stripe 
(tiles) in Fig.~3(a) (Fig.~3(b)) obtained for 2 holes with breathing modes.
However, we can see that the probability of hole 
occupation is almost 1 in the stripe and the tile inhomogeneities 
of Fig.~6 indicating 
localization. In the phase-separated state the holes are localized by 
strong uniform displacements of the oxygens. In the case of the stripe, only 
the oxygens in the links along the stripe are strongly displaced while
the tile is stabilized by a strong displacement of the oxygens 
in the links connecting the holes. In the interval $0.75\le\lambda_0^2/\kappa
\le 2.5$ the tile and the stripe states have very similar energies.
\vskip 2cm

\begin{figure}[thbp]
\begin{center}
\includegraphics[width=6cm,angle=0]{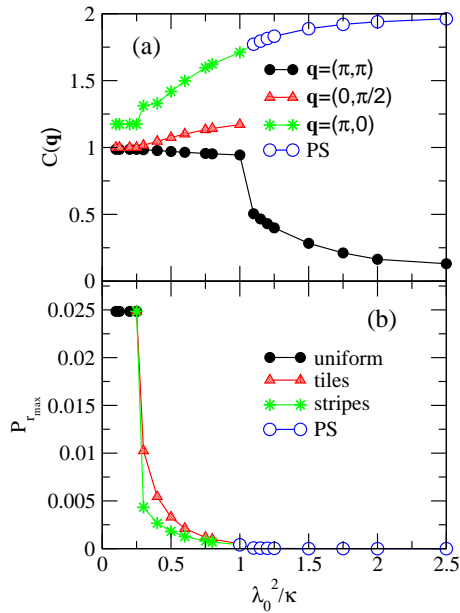}
\vskip 0.3cm
\caption{(a) Charge structure factor as a function of the diagonal 
electron-phonon coupling for the buckling mode and  2 holes in 
a $4\times 4$ lattice; 
(b) Pairing correlations at the maximum distance $r_{max}$ for the same 
parameters as (a)}
\label{fig7}
\end{center}
\end{figure}

At quarter 
filling, i.e., 8 holes, it can be seen that the holes form a cluster so that 
two neighboring rows (or columns) in the $4\times 4$ cluster have electrons 
and the other two rows (or columns) only have holes. Again, the holes 
are localized by strong displacements of the surrounding oxygens. As expected, 
the pairing correlations immediately vanish in the phase separated state.
\vskip 2cm

\begin{figure}[thbp]
\begin{center}
\includegraphics[width=6cm,angle=0]{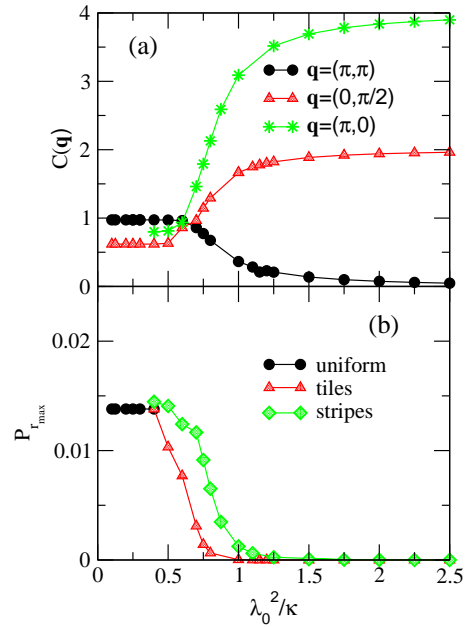}
\vskip 0.3cm
\caption{(a) Charge structure factor as a function of the diagonal 
electron-phonon coupling for the buckling mode and  4 holes in 
a $4\times 4$ lattice; 
(b) Pairing correlations at the maximum distance $r_{max}$ for the same 
parameters as (a).}
\label{fig8}
\end{center}
\end{figure}

Between the uniform and the phase separated states, a region with charge
inhomogeneous structures is observed. 
For two holes we observe in Fig.~7(a) that a peak 
develops in the charge structure factor for $\lambda_0^2/\kappa=0.25$. 
Tiles and stripes similar to those displayed in Fig.~3 become the ground 
state. As the coupling increases the holes become more and more localized in 
the stripe or the tile but phase separation occurs suddenly 
for $\lambda_0^2/\kappa\ge 1$ when the holes form a localized pair. 
In Fig.~7(b) it 
can be seen how the pairing correlations are reduced in the charge 
inhomogeneous phases as localization increases.

A similar behavior, displayed in Fig.~8, 
is observed for 4 holes. The remarkable result is that buckling 
modes appear to be able to stabilize the same
kind of stripes than the half-breathing modes produce, 
but only in a very narrow 
region. 
Let us study in detail what is the underlying ionic 
configuration. In Fig.~9(a) we observe a schematic 
representation of the stripe state
for $\lambda_0^2/\kappa=0.4$. The 
hole density distribution is similar to the one for half-breathing modes 
shown in Fig.~2(b) 
but the ionic displacements are {\bf not} the same. Clearly there is 
a break down of the rotational invariance so we could call the mode 
``half-buckling''. We notice that only the ions in the direction parallel to 
the stripes are displaced, while the breathing modes stabilize the stripe via 
displacements of the ions in the links perpendicular to the stripe. The 
oxygens in the links that connect the hole-rich sites move in one direction 
while those in the links that connect the hole poor sites move in the 
opposite direction. 

This phononic configuration appears as an alternative to the 
half-breathing modes for stabilizing dynamic stripes but it is very unstable.
The induced charge inhomogeneity is very small, of the order of 1\%, and as we
will see below, the 
state is rapidly replaced by a precursor of the phase-separated state.
 
For $\lambda_0^2/\kappa=0.4$ a tile state is also stabilized.
It is shown in Fig.~9(b). We observe that only the oxygens in the links that 
connect the sites that form each tile are displaced. The oxygen displacements 
are opposite in hole rich and hole poor regions. It seems as if a depression 
in the position of the oxygen ions tends to localize the holes, like a 
depression in a mattress would tend to localize marbles. It is interesting 
to monitor how the stripe and tile configurations evolve to the phase 
separated states shown in Fig.~6. At $\lambda_0^2/\kappa=0.5$ 
the stripes become 
distorted. An upward displacement of the oxygen ions along the last column in 
Fig.~10(a) effectively reduces the hole density in that column. A similar 
effect 
is observed in the tile state where, as we can see in Fig.~10(b), an upward 
displacement of the oxygens in one of the tiles pushes the holes out. We have
observed that for $0.4\le\lambda_0^2/\kappa\le 0.5$ the tile state has energy 
slightly lower than the stripe state.

\vskip 2cm

\begin{figure}[thbp]
\begin{center}
\includegraphics[width=6cm,angle=0]{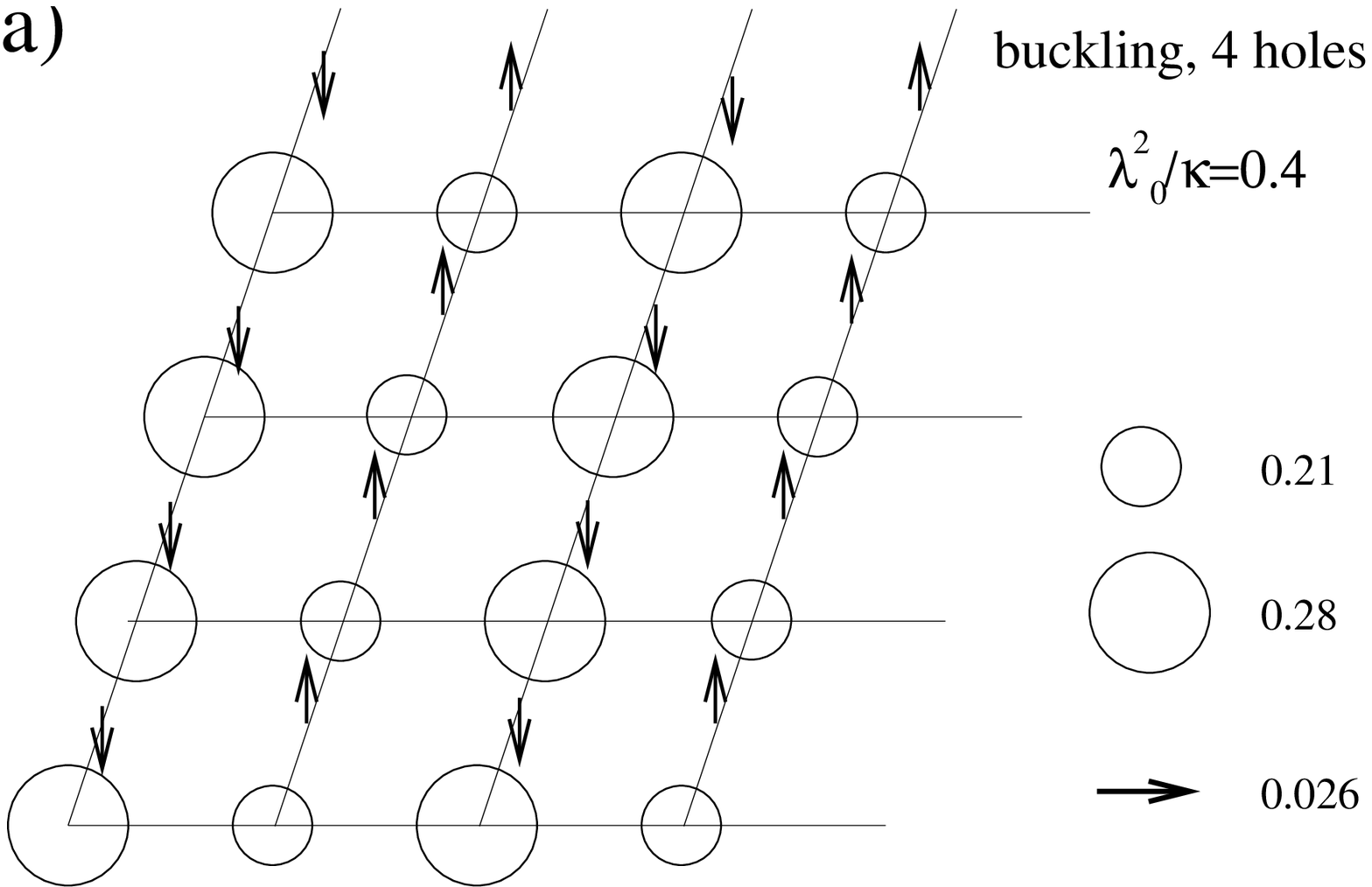}
\vskip 1cm
\includegraphics[width=6cm,angle=0]{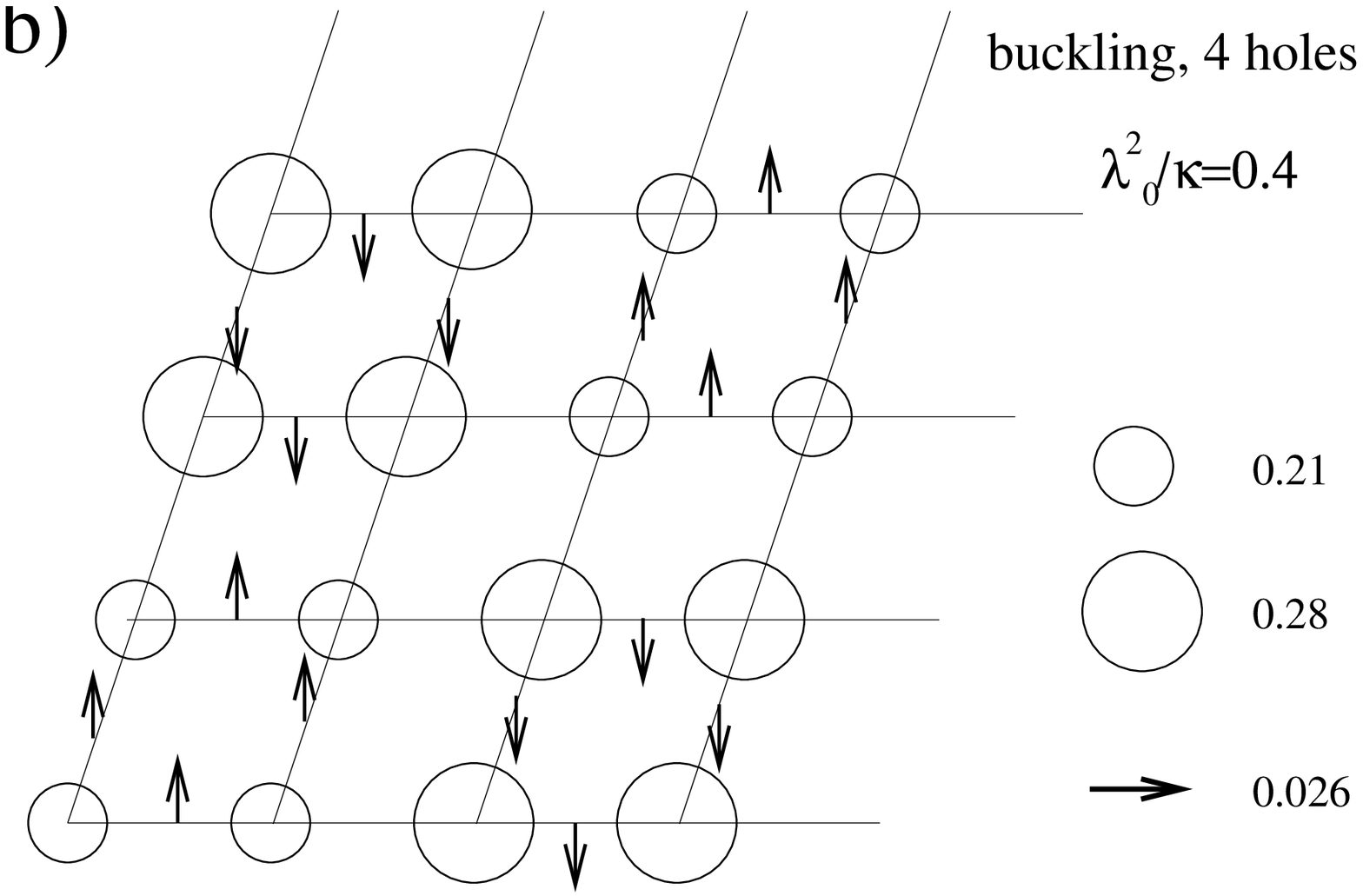}
\vskip 0.3cm
\caption{(a) Schematic representation of the distorted charge 
stripes with weakly localized holes stabilized by buckling phononic modes 
for 4 holes in 
a $4\times 4$ lattice at $\lambda_0^2/\kappa=0.4$; 
(b) Same as (a) but for the tile structure.}
\label{fig9}
\end{center}
\end{figure}
 
\vskip 2cm

\begin{figure}[thbp]
\begin{center}
\includegraphics[width=6cm,angle=0]{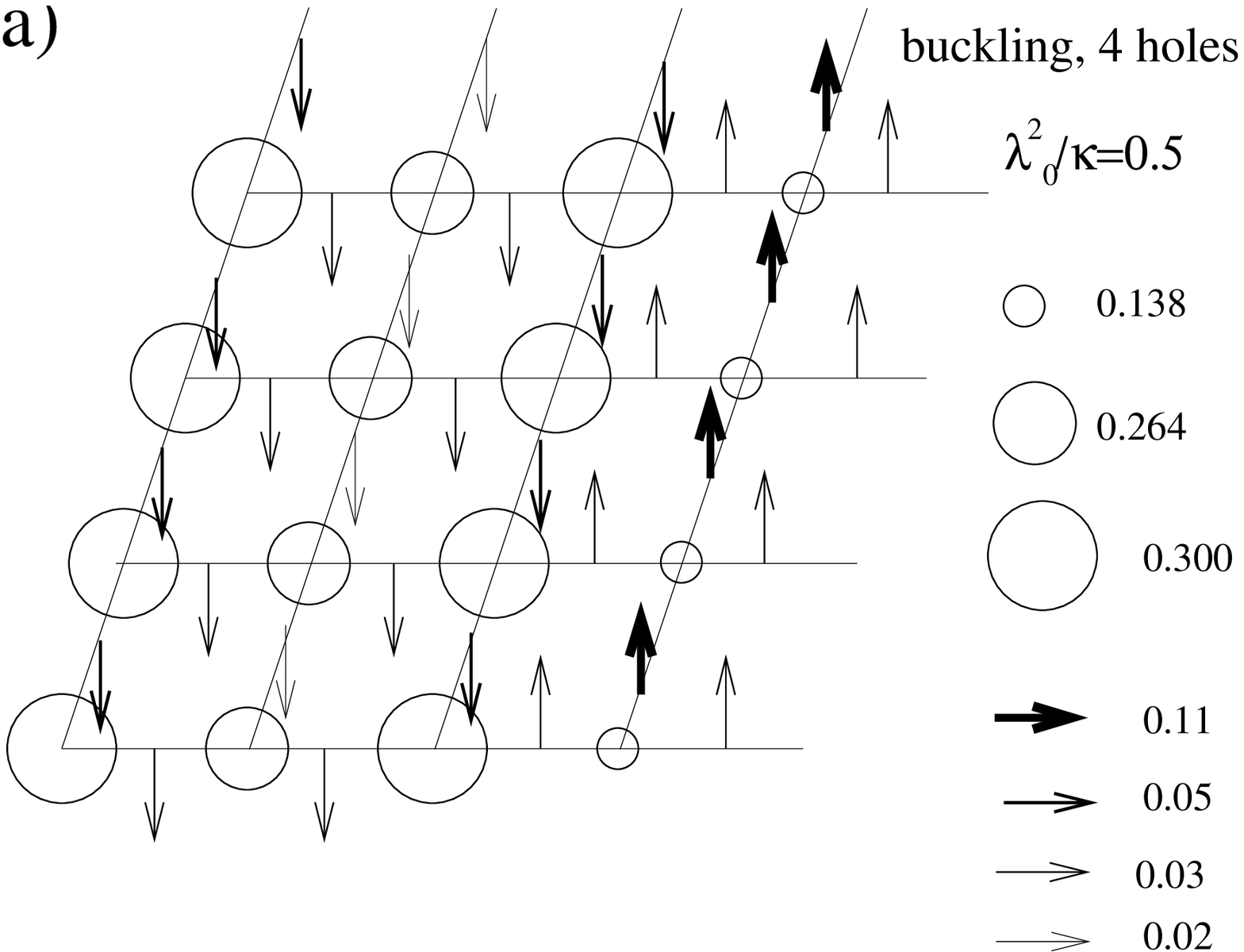}
\vskip 1cm
\includegraphics[width=6cm,angle=0]{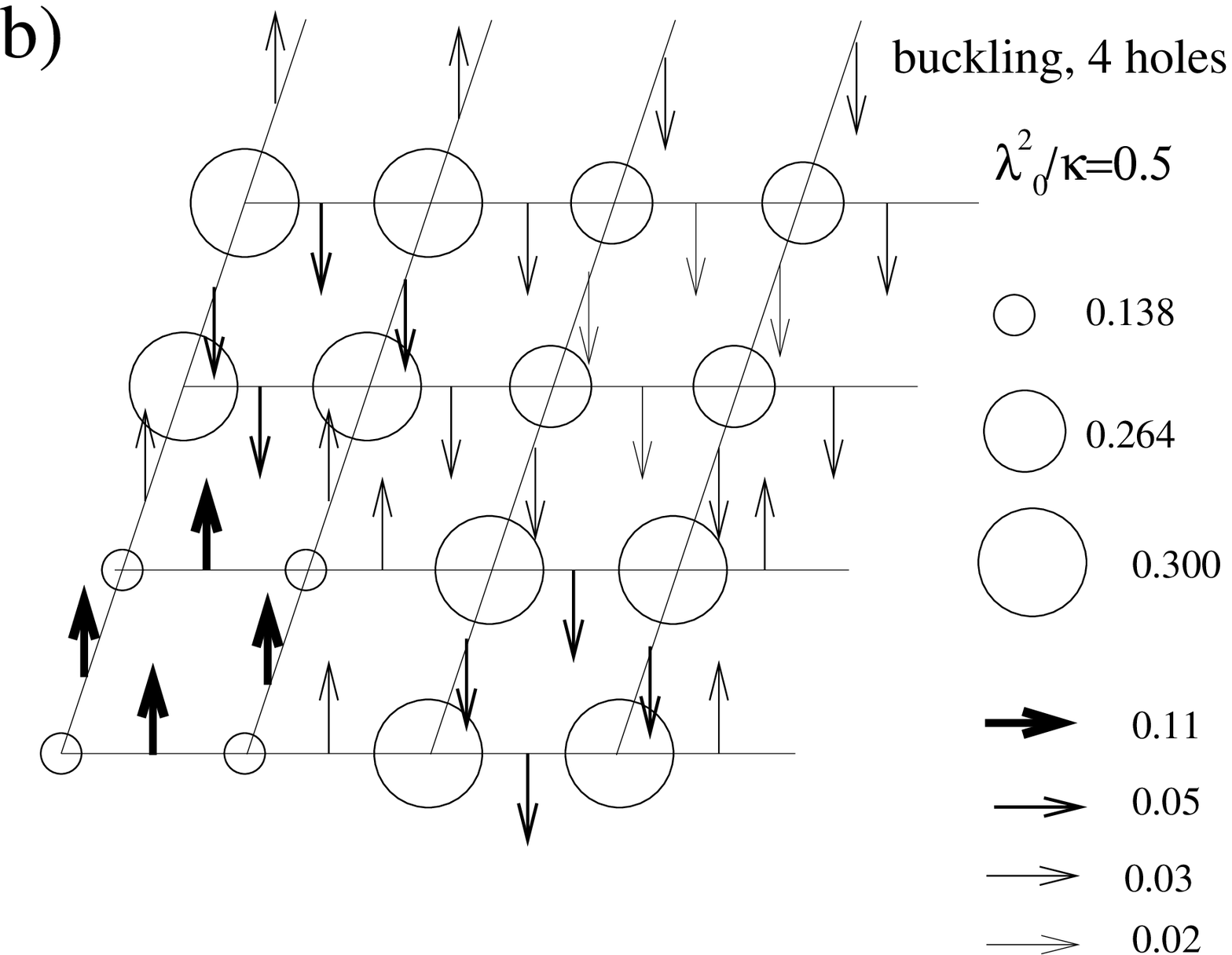}
\vskip 0.3cm
\caption{(a) Schematic representation of the charge 
stripes stabilized by buckling phononic modes for 4 holes in 
a $4\times 4$ lattice; 
(b) Same as (a) but for the tile structure stabilized by 
buckling phononic modes.}
\label{fig10}
\end{center}
\end{figure}
 
An important qualitative 
change occurs at $\lambda_0^2/\kappa=0.6$. Both in the stripe 
(Fig.~11(a)) and in the tile (Fig.~11(b)) 
configurations the holes get localized in 
the region of the lattice in which the oxygens are depressed. The stripe 
state has the lowest energy in this case.

\vskip 2cm

\begin{figure}[thbp]
\begin{center}
\includegraphics[width=6cm,angle=0]{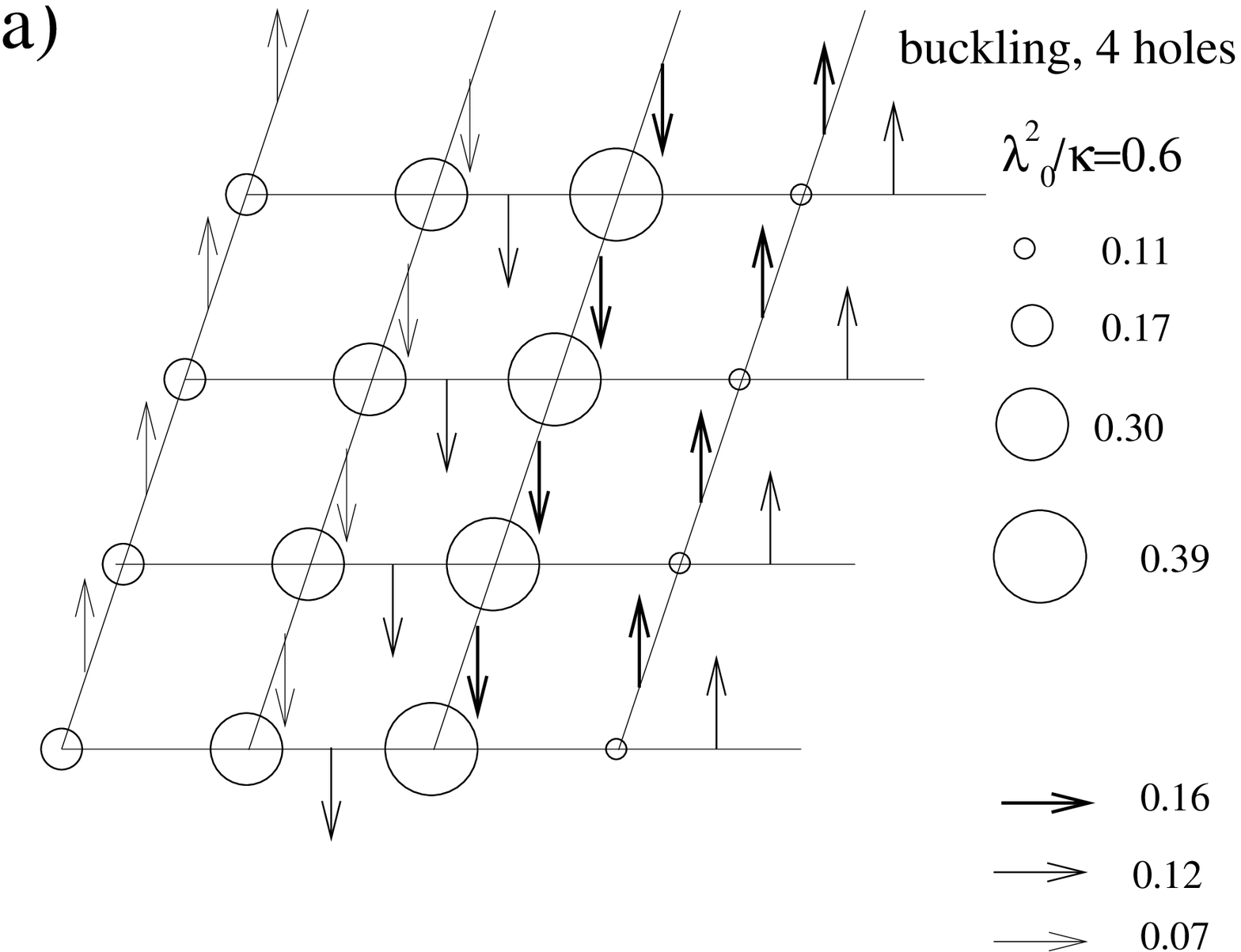}
\vskip 1cm
\includegraphics[width=6cm,angle=0]{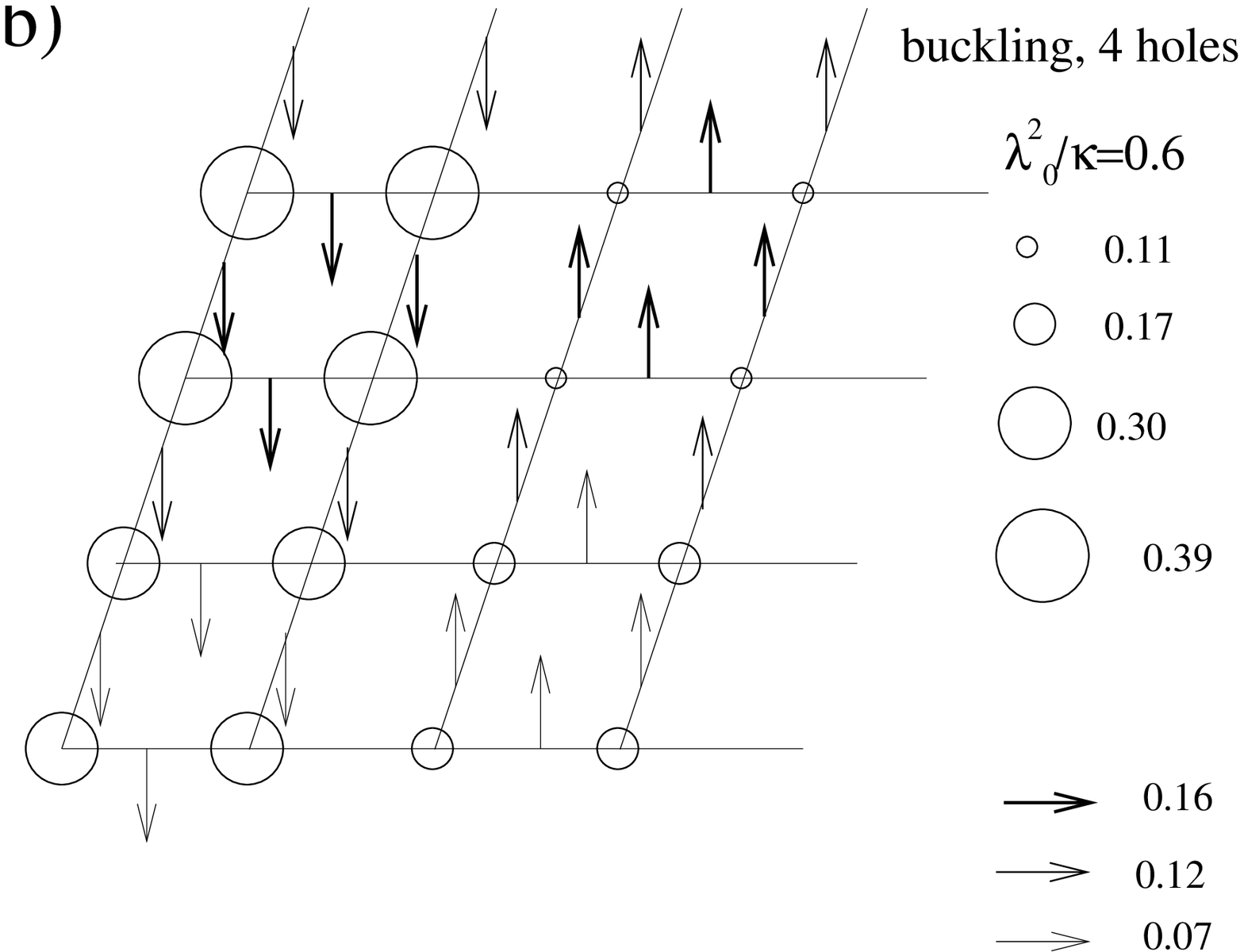}
\vskip 0.3cm
\caption{(a) Schematic representation of the charge 
stripes, precursor of the phase separated state, 
stabilized by buckling phononic modes for 4 holes in 
a $4\times 4$ lattice for $\lambda_0^2/\kappa=0.6$; 
(b) Same as (a) but for the tile structure.}
\label{fig11}
\end{center}
\end{figure}
 
Finally, at $\lambda_0^2/\kappa=0.7$ a structure similar to the one shown 
in Fig.~6 
develops. The ion displacements become stronger as the diagonal coupling 
increases as well as the local differences in hole density. Actual phase 
separation, with $n(i)\approx 0$ or $1$ in every site, i.e., 
with the holes completely localized, appears to develop for 
$\lambda_0^2/\kappa\ge 1$.  

In Fig.~8(b) we can see the behavior of the pairing correlations in the different 
phases. It is interesting to compare with the corresponding results in Fig.~1(b)
for the 
half-breathing case. The pairing correlations in the uniform state are, of 
course, the same as in Fig.~1(b). Thus, to simplify the figures we only present 
results for the state with momentum $(0,\pi)$ in the uniform case despite the 
fact that we know that the ground state is triply degenerate in momentum. 
The filled circles indicate the pairing along the maximum diagonal distance. 

When the double stripe is stabilized for $\lambda_0^2/\kappa=0.4$ we observe
a behavior very similar to the one for stripes in the breathing case. However, 
the stripes are destabilized by a phase which is a precursor of the phase 
separated state where the holes are localized by the 
buckling modes (see Fig.~10). 
The area of the localization region decreases with the 
diagonal EPI until the holes form a localized cluster and phase separation 
is reached. The pairing correlations, as expected, decrease sharply.

An interesting detail is that, as in the half-breathing case, 
in all stripe states the pairing correlation 
is larger in the 
direction perpendicular to the stripes that in the direction parallel to them.

At quarter filling the transition from the uniform to the phase-separated 
state is 
more abrupt. In fact, only for
$\lambda_0^2/\kappa=0.9$ the localization region is larger than two rows 
or columns.
In Fig.~12(a) it is shown how the weak peak at $(\pi,\pi)$ in the charge 
structure factor characteristic of the uniform ground state is replaced 
by a strong peak at $(0,\pi/2)$ indicating that half the lattice is filled 
with holes and the other half is empty. On the other hand, the magnetic 
incommensurability characteristic of the uniform quarter-filled state 
dissapears in the phase separated region as it can be seen in Fig.~12(b).

\vskip 2cm
\begin{figure}[thbp]
\begin{center}
\includegraphics[width=6cm,angle=0]{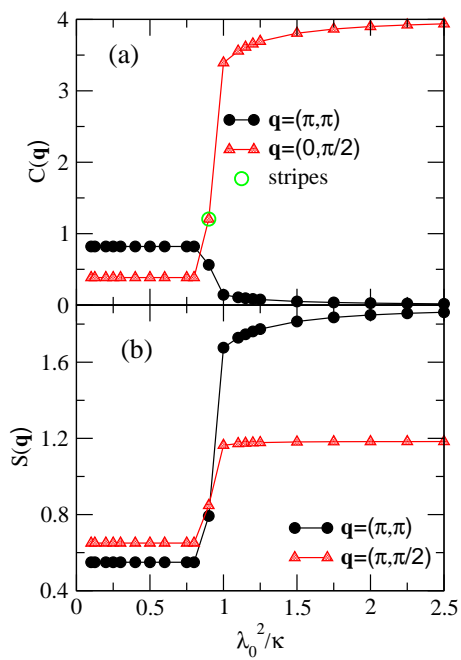}
\vskip 0.3cm
\caption{(a) Charge structure factor as a function of the diagonal 
electron-phonon coupling for the buckling mode and 8 holes in 
a $4\times 4$ lattice; 
(b) Magnetic structure factor for the same 
parameters as (a).}
\label{fig12}
\end{center}
\end{figure}

Summarizing, the buckling modes induce a lattice depression where the holes
concentrate. A ``half-buckling'' mode that plays a role similar to the 
``half-breathing'' mode is rapidly destabilized by the diagonal coupling in
favor of a state with localized clustered holes and suppressed pairing 
correlations, which is a precursor of the phase-separated state.

\subsection{Off-Diagonal Electron-Phonon coupling}

At the adiabatic level, we have observed that
the effect of the off-diagonal terms is to 
destabilize the charge inhomogeneous states with mobile holes that appear in 
between the uniform
and the localized holes or phase separated phases. 
This can be seen in Fig.~13 where 
a phase diagram in the plane $\lambda_t$ vs $\lambda_0^2/\kappa$ for breathing 
modes is presented 
for 4 holes and $\lambda_J=\lambda_t$. 
It can be seen that for 
$\lambda_0^2/\kappa\approx 0.7$ the 
charge inhomogeneous stripe/tile state is replaced by an uniform ground state 
for $\lambda_t=0.1$. As $\lambda_0$ increases, larger values of $\lambda_t$ 
are needed to restore the uniform ground state. The region labeled $M$ in the 
figure is characterized by charge inhomogeneous states with mobile holes. 
This phase is overcome by the uniform state for larger values of the 
off-diagonal coupling. We have not observed any enhancement of the pairing 
correlations due to the off-diagonal coupling.

For buckling modes, see Fig.~14, we observe that small values of $\lambda_t$
just renormalize the diagonal coupling to a smaller value so that the uniform 
ground state remains stable up to larger values of $\lambda_0^2/\kappa$. 
However, larger values of $\lambda_t$ destabilize the inhomogeneous phase 
with mobile holes  
labeled $M$ in the figure, and for $\lambda_t>0.2$ the system goes from a 
uniform to a phase separated state as a function of $\lambda_0^2/\kappa$. 

The off-diagonal coupling does not generate new phases, it simply seems to 
renormalize the diagonal coupling. In fact, its effect 
is to destabilize the inhomogeneous phase with mobile holes identified by 
$M$ in Figs.~13 and 14. This phase is more easily destabilized for buckling 
than for breathing modes.

\vskip 2cm
\begin{figure}[thbp]
\begin{center}
\includegraphics[width=6cm,angle=0]{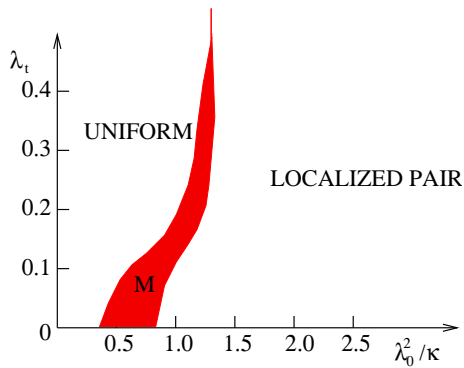}
\vskip 0.3cm
\caption{Phase diagram for 4 holes as a function of the diagonal EPI 
$\lambda_0^2/\kappa$ and the off-diagonal coupling $\lambda_t$ 
for breathing modes. In this case $\lambda_J=\lambda_t$. {\it M} characterizes 
charge inhomogeneous states with mobile holes described in the text.}
\label{fig13}
\end{center}
\end{figure}

\vskip 2cm
\begin{figure}[thbp]
\begin{center}
\includegraphics[width=6cm,angle=0]{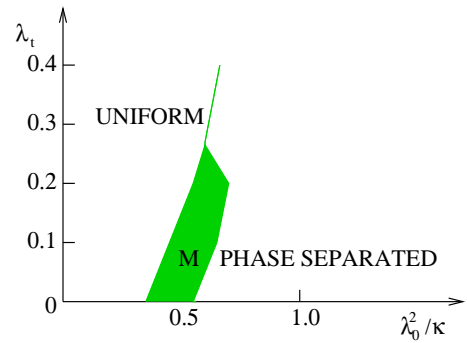}
\vskip 0.3cm
\caption{Phase diagram for 4 holes as a function of the diagonal EPI 
$\lambda_0^2/\kappa$ and the off-diagonal coupling $\lambda_t$ for buckling 
modes. 
In this case $\lambda_J=\lambda_t$. {\it M} characterizes 
charge inhomogeneous states with mobile holes described in the text.}
\label{fig14}
\end{center}
\end{figure}

\section{Quantum Phonons}

The results presented in the previous section indicate that the phonon mode 
that most closely reproduces the experimental results observed in the cuprates 
is the half-breathing. In addition, neutron scattering experiments 
show that this is the mode the most strongly coupled to the charge in the 
high $T_c$ cuprates. For this reason we decided to study the effects of the 
half-breathing mode beyond the adiabatic approximation.

In order to study the electron-phonon interaction at finite frequency
the electron-phonon part of the Hamiltonian (Eq.(1b)) should be 
replaced by\cite{doug} 
$$
H_{e-ph}=\lambda_0\sum_{{\bf i,\mu}}(b^{\dagger}_{{\bf i,\mu}}+ 
b_{{\bf i,\mu}})(n_{{\bf i}}\mp n_{{\bf i+\mu}}),
\eqno(8a)
$$
\noindent where the minus sign between the densities corresponds to the 
breathing mode while the plus sign corresponds 
to the buckling mode that will be 
discussed later. 
$b^{\dagger}_{{\bf i,\mu}}$ creates a 
phonon at the link ${\bf i,\mu}$ with frequency $\omega$ and 
$\lambda_0=g\sqrt{{1\over{2m\omega}}}$, where $g$ is the electron-phonon 
coupling and $m$ is the ionic mass. Eq.(1c) should be replaced by
$$
H_{ph}=\omega\sum_{{\bf i,\mu}}(b^{\dagger}_{{\bf i,\mu}}b_{{\bf i,\mu}}
+{1\over{2}}).
\eqno(8b)
$$
The off-diagonal effects are given by
$$
t_{{\bf i,j}}=t\{1+\lambda_t[\hat B({\bf i})+\hat B({\bf j})]\},
\eqno(9)
$$
\noindent and
 $$
J_{{\bf i,j}}=J[1+\lambda_J[\hat B({\bf i})+\hat B({\bf j})]\}, 
\eqno(10)
$$
\noindent where
$$
\hat B(i)=b_{{\bf i,x}}+
b^{\dagger}_{{\bf i,x}}-b_{{\bf i-\hat x,x}}-
b^{\dagger}_{{\bf i-\hat x,x}}+b_{{\bf i,y}}+
b^{\dagger}_{{\bf i,y}}-b_{{\bf i-\hat y,y}}-
b^{\dagger}_{{\bf i-\hat y,y}}.
\eqno(11)
$$
In order to simplify these expressions we make the Fourier transform of 
phonon and electron operators. Then, in momentum space, the interaction 
term becomes:
$$
H_{e-ph}={1\over{\sqrt{N}}}\sum_{{\bf k,q},\mu,\sigma}\lambda_0
(1\mp e^{-i{\bf q\cdot\hat\mu}})
c^{\dagger}_{{\bf k},\sigma}c_{{\bf k-q},\sigma}(b^{\dagger}_{{\bf q},\mu}+
b_{{\bf -q},\mu}).
\eqno(12)
$$
The effective 
electron-phonon coupling constant $\lambda({\bf q},\mu)=\lambda_0
(1\mp e^{-i{\bf q\cdot\hat\mu}})$ is clearly anisotropic. 
For breathing modes (negative sign inside the parenthesis) it is 
strongest for ${\bf q}_{max}=(\pi,\pi)$ and, thus, in the simplest 
approximation only the phonon
mode $b_{{\bf q}_{max}}$ is maintained.\cite{doug,jose} 
For half-breathing modes 
${\bf q}_{max}=(\pi,0)$ or $(0,\pi)$. Under this approximation the Hamiltonian 
for half-breathing modes along $x$ becomes:
$$
H_{e-ph}^{br}=2\lambda_0(b^{\dagger}+b)\sum_{\bf i}n_{\bf i}(-1)^{x_{\bf i}},
\eqno(13)
$$
\noindent where $x_{\bf i}$ is the x-coordinate of site ${\bf i}$.
$$
H_{ph}^{br}=\omega(b^{\dagger}b+{1\over{2}}).
\eqno(14)
$$
The off-diagonal effects on the hopping and the Heisenberg 
coupling vanish along the direction of the half-breathing modes 
($\hat x$ in this case). Along the perpendicular direction they are given 
by
$$
t^{br}_{{\bf i,i+\hat y}}=t[1+4\lambda_t (-1)^{x_i} (b^{\dagger}+b)],
\eqno(15)
$$
\noindent and
$$
J^{br}_{{\bf i,i+\hat y}}=J[1+4\lambda_J (-1)^{x_i} (b^{\dagger}+b)].
\eqno(16)
$$
This behavior was also remarked in Ref.\onlinecite{ishihara}.
\vskip 2cm
\begin{figure}[thbp]
\begin{center}
\includegraphics[width=6cm,angle=0]{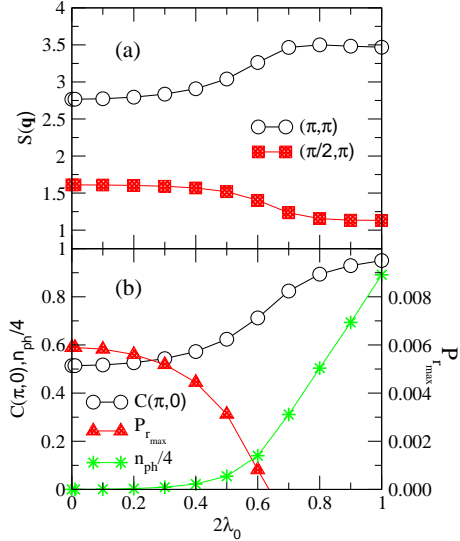}
\vskip 0.3cm
\caption{(a) Magnetic structure factor for different values of the momentum 
as a function of the diagonal electron-phonon coupling, for quantum 
half-breathing 
modes in a $4\times 4$ lattice for 2 holes.
(b) Charge structure factor, pairing at maximum distance, and average number 
of phonons as a function of the diagonal electron-phonon coupling, for quantum 
half-breathing modes in a $4\times 4$ lattice with 2 holes.}
\label{fig15}
\end{center}
\end{figure}

\vskip2cm
\begin{figure}[thbp]
\begin{center}
\includegraphics[width=6cm,angle=0]{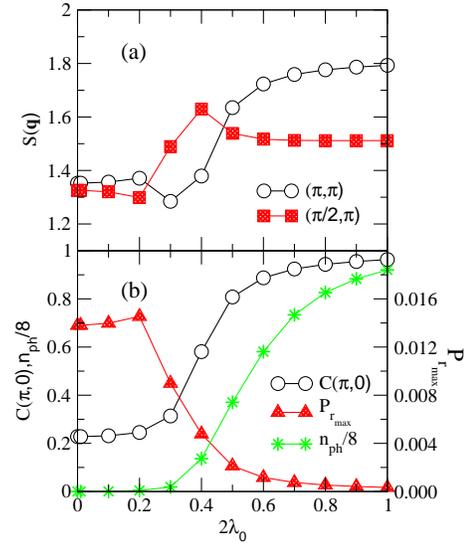}
\vskip 0.3cm
\caption{(a) Magnetic structure factor for different values of the momentum 
as a function of the diagonal electron-phonon coupling, for quantum 
half-breathing 
modes in a $4\times 4$ lattice for 4 holes.
(b) Charge structure factor, pairing at maximum distance and average number 
of phonons as a function of the diagonal electron-phonon coupling, for quantum 
half-breathing modes in a $4\times 4$ lattice for 4 holes.}
\label{fig16}
\end{center}
\end{figure}

\begin{figure}[thbp]
\begin{center}
\includegraphics[width=6cm,angle=0]{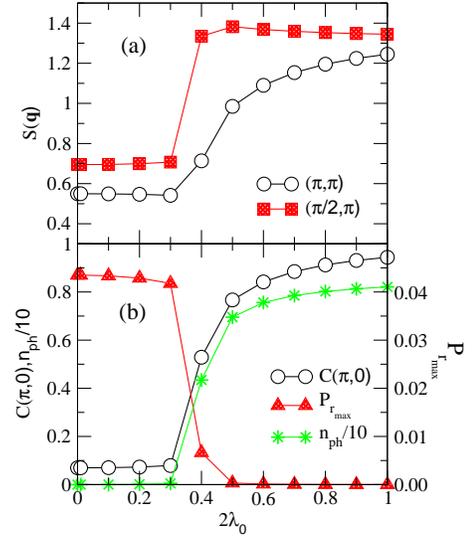}
\vskip 0.3cm
\caption{(a) Magnetic structure factor for different values of the momentum 
as a function of the diagonal electron-phonon coupling for 
quantum half-breathing 
modes in a $4\times 4$ lattice for 8 holes.
(b) Charge structure factor, pairing at maximum distance and average number 
of phonons as a function of the diagonal electron-phonon coupling for quantum 
half-breathing modes in a $4\times 4$ lattice for 8 holes.}
\label{fig17}
\end{center}
\end{figure}

As in the previous sections, 
$J$ will be fixed to 0.4; $\omega$ will be set to 1.

Let us discuss first the effects of the diagonal coupling, i.e., 
$\lambda_t=\lambda_J=0$. An important difference between the adiabatic and the 
quantum case is that with adiabatic phonons the ionic displacements were 
unconstrained and, as a result, we observed that in some regions of parameter 
space extended breathing modes with rotational symmetry that stabilized tile, 
CDW, and localized holes structures, were energetically more favorable than 
half-breathing modes. With quantum phonons, we will study half-breathing modes 
only and, thus, all the charge inhomogeneous states that will arise will be 
stripe-like. 

In Fig.~15 results for 2 holes doped in a $4\times 4$ lattice are 
presented. In panel (a) we observe the magnetic structure factor for different 
values of the momentum. It can be seen that the value at $(\pi,\pi)$ is always 
maximum and, as a result, magnetic incommensurability does not develop. 
The charge structure factor (open circles in panel (b)) indicates 
that a stripe-like structure starts to develop at $2\lambda_0=0.5$, i.e., 
when the average number of phonons
(star symbols) becomes different from zero. 
However, the pairing correlations at the maximum diagonal distance 
(indicated by triangles) always decrease for finite EPI and actually vanish 
in the stripe phase. This occurs because the stripe pattern that results in 
this case is different from the physically ``correct" stripe shown in Fig.~3(b)
that 
we obtained in the adiabatic case. In order to obtain this pattern with 
quantum phonons, finite values of $g(\pi/2,0)$ should be consider in addition 
to finite values of $g(\pi,0)$, as explained in Section II. 
The single coupling at momentum $(\pi,0)$ 
stabilizes a double stripe pattern with one hole per stripe. The charge 
difference between the background and stripes is large enough to develop 
a peak at $(\pi,0)$, but the behavior of the magnetic
structure factor indicates that it does not produce a $\pi$-shift.

In Fig.~16, results for different values of $\lambda_0$  on a 
$4\times 4$ lattice doped with four holes are presented. In this case we 
obtain 
results qualitatively very similar to those in the adiabatic case 
(Fig.~1). The magnetic 
structure factor displayed in Fig.~16(a) develops a peak at ${\bf q}=(\pi/2,\pi)$
for $0.2<2\lambda_0<0.5$, which indicates that a state with magnetic 
incommensurability is stabilized. The charge structure factor shown in 
Fig.~16(b)
(with open circles) indicates that charge order consistent with two vertical 
stripes develops in the magnetically incommensurate region. 
The uniform ground state is replaced by an inhomogeneous one when the number 
of phonons (star symbols) becomes finite at $2\lambda_0>0.2$. This is similar 
to the effect observed with adiabatic phonons.
For $2\lambda_0>0.5$ the charge structure
factor results indicates that a static stripe-like structure has developed. A
$\pi$-shift does not occur since $S({\bf q})$ has a maximum at $(\pi,\pi)$. 

A small  enhancement in the pairing correlations at maximum diagonal distance 
(triangles in Fig.~16(b)) occurs as a function of $\lambda_0$ in the uniform 
phase. As in the adiabatic case, D-wave pairing correlations are {\bf not} 
dramatically affected in charge inhomogeneous states with mobile holes.  


Fig.~17 shows the results at quarter-filling. Instead of a CDW stabilized by 
full-breathing modes, which is what we observed in the adiabatic case, 
the half-breathing modes still generate a two-stripe structure. This occurs
because we are still studying the effects of half-breathing modes, while phonon
operators with momentum $(\pi,\pi)$ would induce the CDW state. 
The strong peak in the charge structure factor $C$ at $(\pi,0)$ 
(open circles in Fig.~17(b)) underlines the large 
charge inhomogeneity which induces magnetic incommensurability, as it can be 
seen from the magnetic structure factor's behavior displayed in Fig.~17(a).
\vskip 2cm
\begin{figure}[thbp]
\begin{center}
\includegraphics[width=6cm,angle=0]{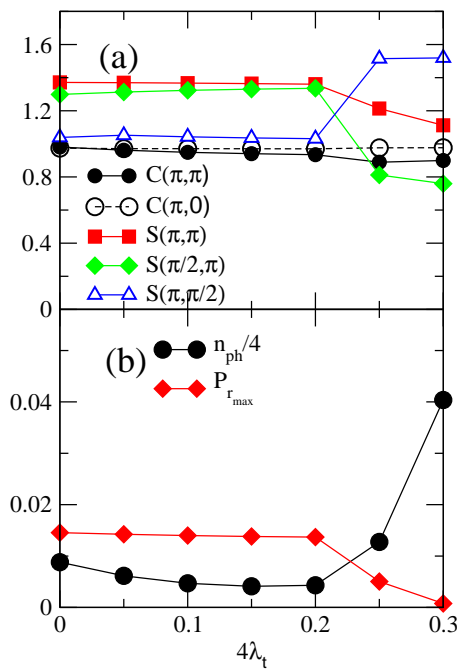}
\vskip 0.3cm
\caption{(a) Charge (C) and Magnetic (S) 
structure factors for different values of the 
momentum, as a function of the off-diagonal electron-phonon coupling 
$\lambda_t$ for
quantum half-breathing 
modes in a $4\times 4$ lattice for 4 holes. $\lambda_J=2\lambda_t$ and the 
diagonal coupling is given by $2\lambda_0=0.2$.
(b) Pairing at maximum diagonal distance, 
and average number 
of phonons as a function of the 
off-diagonal electron-phonon coupling for quantum 
half-breathing modes for the same parameters as in part (a).}
\label{fig18}
\end{center}
\end{figure}

The tendency to incommensurate order increases as the doping is increased from
2 to 8 holes, as it can be seen in Figs.~15, 16 y 17 ((a) panels). This is 
understandable since the antiferromagnetic ordering between the two undoped
domains outside the stripes is reduced when the stripes become increasingly
populated by holes.

Let us now consider the effects of off-diagonal electron-phonon coupling. 
We will start with the case of 4 holes at $2\lambda_0=0.2$ which, 
according to Fig.~16(b), provides the state with the strongest pairing 
correlations. We have set $\lambda_J=2\lambda_t$.\cite{ishihara} The state at
$\lambda_t=0$ is uniform and, like in the adiabatic case (Fig.~1), is triply 
degenerate in momentum. The off-diagonal EPI breaks the degeneracy in 
momentum and selects de ground state with momentum $(0,\pi)$. As $\lambda_t$ 
increases the peak in the charge structure factor at momentum $(\pi,0)$ 
remains almost unchanged (circles in Fig.~18(a)). The peaks at 
momentum $(\pi,\pi)$ (squares) and $(\pi/2,\pi)$ (diamonds) 
in the magnetic structure factor remains essentially unchanged as well.
While this happens the pairing correlations and the 
average number of phonons decrease slightly (Fig.~18(b)). 
This trend continues up to $4\lambda_t=0.2$. Beyond this 
value, there are no indications of the formation of a charge inhomogeneity
but there are some changes in the magnetic channel with an incipient order with
momentum $(\pi,\pi/2)$.

We will also study the effects of off-diagonal interactions for 
$2\lambda_0=0.4$, i.e., in the region with dynamical stripes. The effect of 
$\lambda_t$ in this case is to select the state with momentum (0,0) as the 
ground state. At $\lambda_t=0$  the ground state presents a charge 
inhomogeneous state with two stripes according to the large peak at $(\pi,0)$
in $C({\bf q})$ (circles in Fig.~19(a)) 
exhibiting magnetic incommensurability. 
The pairing is very 
reduced. As $\lambda_t$ increases up to 
$4\lambda_t=0.1$ the charge inhomogeneity is reduced and the pairing 
correlations increase indicating that $\lambda_t$ is renormalizing $\lambda_0$
making it effectively smaller and stabilizing the charge uniform state.
This is clearly seen at $4\lambda_t=0.15$ when the ground state changes 
momentum
to $(\pi,0)$ and we observed a ground state with properties similar, 
including the pairing, to the case $\lambda_0=0.2$ and $\lambda_t=0$.  
The off-diagonal coupling just seems to make the diagonal one weaker.
\vskip 2cm
\begin{figure}[thbp]
\begin{center}
\includegraphics[width=6cm,angle=0]{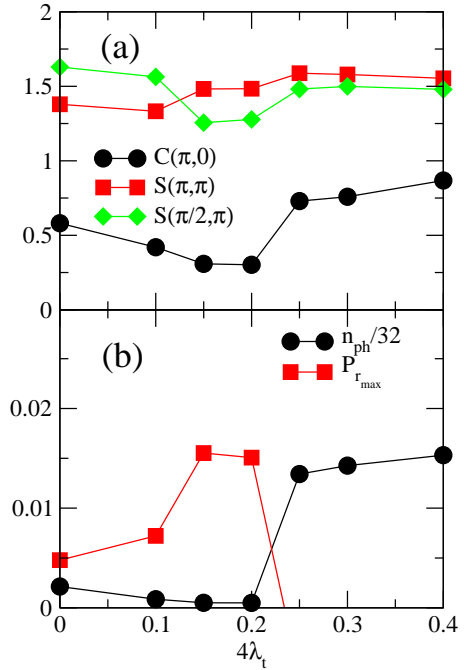}
\vskip 0.3cm
\caption{(a) Charge and magnetic structure factor for different values of the 
momentum as a function of the off-diagonal electron-phonon coupling 
$\lambda_t$ for
quantum half-breathing 
modes in a $4\times 4$ lattice for 4 holes. $\lambda_J=2\lambda_t$ and the 
diagonal coupling is given by $2\lambda_0=0.4$.
(b) Pairing at maximum diagonal distance, 
and average number 
of phonons as a function of the 
off-diagonal electron-phonon coupling for quantum 
half-breathing modes for the same parameters as in part (a).}
\label{fig19}
\end{center}
\end{figure}

This level crossing is related to the disappearance of the magnetic 
incommensurability at $(\pi/2,\pi)$. Once this incommensurability has
disappeared, a second level crossing takes place at $4\lambda_t \approx 0.225$
and the ground state momentum is again $(0,0)$. By further increasing 
$\lambda_t$, the charge ordering is reestablished. This complex behavior,
together with the one previously found for $2\lambda_0=0.2$, 
illustrates the interplay between the diagonal and off-diagonal EPI. In
particular, due to a nonzero $\lambda_J$, the spin degrees of freedom are now
directly coupled to the ion displacements, and hence magnetic orderings are
more relevant to the overall behavior than with purely diagonal EPI.

For completeness we are going to comment on the effects of the diagonal 
electron-phonon coupling to buckling modes with quantum phonons. 

Our studies with adiabatic phonons showed that in a very narrow region of 
parameter space stripes were stabilized by a buckling mode that breaks 
rotational invariance. The mode that stabilizes vertical stripes would have a 
maximum coupling constant at momentum $(\pi,0)$. Selecting this momentum in
Eq.(12) (with the plus sign) we arrive to the same Hamiltonian as in 
the half-breathing case, except that the single phonon modes are now along 
the $y$ direction (parallel to the stripe) rather that in the direction $x$ as
in the half-breathing case, in agreement with the results obtained in the 
adiabatic case (Fig.~9(a)). In addition, off-diagonal effects vanish in 
this case. Thus, the same physical properties as in the diagonal 
half-breathing case are obtained.

In Ref.~\onlinecite{nazarenko} it was shown at the mean-field level 
that a diagonal 
coupling was sufficient to enhance D-wave pairing correlations. Selecting the 
plus sign in Eq.(12) we observe that the effective coupling constant is 
maximum for ${\bf q}=(0,0)$ and vanishes at momentum $(\pi,\pi)$
in agreement with the result of Nazarenko et al.\cite{nazarenko}. 
Regretfully keeping only momentum ${\bf q}=(0,0)$ in Eq.(12) 
leads to
$$
H_{e-ph}^b=2\lambda_0\sum_{\mu}(b^{\dagger}_{\mu}+b_{\mu})\sum_{\bf i}
n_{\bf i},
\eqno(17)
$$
\noindent and
$$
H_{ph}^b=\omega\sum_{\mu}(b^{\dagger}_{\mu}b_{\mu}+{1\over{2}}),
\eqno(18)
$$
\noindent which does not couple the electrons to the phonons since the number 
of electrons is fixed in our simulations and $\sum_{\bf i}n_{\bf i}$ is 
constant for all states. A study of this case conserving more than 
two phonon modes 
would be desirable since mean field calculations \cite{nazarenko} and 
results for 2 holes doping obtained with a truncated approach\cite{doug} 
provided encouraging results towards D-wave pairing stabilization.

\section{Conclusions}

Summarizing, we have studied the effects of electron-phonon interactions in 
the t-J model for a variety of phononic modes, couplings, and electronic 
densities using small clusters solved exactly. 
The most remarkable result is that charge inhomogeneous states with 
mobile holes are stabilized by these interactions. In particular, stripe states
which are very difficult to stabilize in the t-J model are clearly observed.
Tile structures, detected experimentally in some cuprates, are stabilized for 
the first time in the t-J model. Our calculations also confirm that the 
half-breathing mode that stabilizes the stripes is the most energetically 
favorable in some regions of parameter space. We have also found that 
buckling modes can generate stripe and tile states in a very narrow range of
electron-phonon coupling constants. Thus, this mode also has to be considered 
in the interpretation of experimental data. The pairing correlations vanish
in the regime of localized holes induced by the EPI but it is encouraging 
that not drastic effects occur in the states with mobile holes. In fact, 
small enhancements have been detected in some cases. This should be contrasted
with the pair breaking effects of the full breathing mode in D-wave 
superconductors. As previously observed 
in the context of the spin-fermion model \cite{yucel} and diagramatic 
calculations, \cite{naga} diagonal EPI induce 
charge localization while off-diagonal EPI encourages hole mobility.

Our results indicate that electron-phonon interactions need to be considered 
in order to understand the properties of the high-Tc cuprates, in particular, 
the existence of charge inhomogeneous ground states.

\section{Acknowledgments} 

We acknowledge discussions with T. Egami, J. Tranquada, G. Sawatzky and E.
Dagotto. A.M. is supported by NSF under grants DMR-0443144 and DMR-0454504.
Additional support is provided by ORNL.


\begin{thebibliography}{}
\bibitem{foot1} The pairing correlations presented in this paper are D-wave
which are the most enhanced in the t-J model without phonons. We have also
studied extended-S and P-wave pairings but these correlations are always 
weaker than the D-wave ones.
\end{thebibliography}
\end{document}